\documentclass[
aip,
jap,
preprint,
tightenlines,
amsfonts,
amsmath,
amssymb,
nofootinbib]
{revtex4-1}
\usepackage[utf8]{inputenc}
\usepackage[T2A]{fontenc}
\usepackage{hyphenat}
\usepackage{indentfirst}
\usepackage{graphicx}
\usepackage{float}
\usepackage{dcolumn}
\usepackage{multirow}
\usepackage{booktabs}
\usepackage{bm}
\usepackage{color}
\usepackage{hyperref}
\usepackage{xcolor}
\usepackage{epsfig}
\usepackage{epstopdf}

\newcommand{\ddtp}[1]{\frac{\partial{#1}}{\partial{t}}}
\newcommand{\dif}{\mathrm{d}}

\newcommand{\const}{\mathrm{const}}

\begin{document}
\title{Analysis of sample temperature dynamics under pulsed laser irradiation during laser-induced-desorption diagnostic}
\author{A.A. Stepanenko}
\email{aastepanenko@mephi.ru}
\affiliation{National Research Nuclear University MEPhI (Moscow Engineering Physics Institute), 115409, Moscow, Kashirskoe highway, 31, Russia}
\author{Yu.M. Gasparyan}
\email{ymgasparyan@mephi.ru}
\affiliation{National Research Nuclear University MEPhI (Moscow Engineering Physics Institute), 115409, Moscow, Kashirskoe highway, 31, Russia}
\date{\today}
\pacs{}
\keywords{}
\begin{abstract}
The accurate assessment of the local tritium concentration in the tokamak first wall by means of the laser-induced desorption (LID) diagnostic is sought as one the key solutions to monitoring the local radioactive tritium content in the first wall of the fusion reactor ITER. Numerical models of gas desorption from solids used for LID simulation are usually closed with the one-dimensional heat transport models. In this study, the temperature dynamics of a target irradiated by a short laser pulse during LID are analyzed by means of the two-dimensional heat transport model to assess the validity of using one-dimensional approximation for recovering the diagnostic signal. The quantitative estimates for the parameters governing the heat transfer are presented. The analytical expressions for the sample temperature distribution resolved both in time and space are derived. The sensitivity analysis of the obtained relations to uncertainties in the experimental parameters is performed. It is shown that, depending of the ratio between the laser spot radius and heat diffusion length, the one-dimensional approach can noticeably overestimate the sample temperature in the limit of small laser spot radius, resulting in more than 100\% larger amounts of tritium desorbed from the irradiated target, compared to the two-dimensional approximation. In the limit of large laser spot radius, both approaches yield comparable amounts of desorbed tritium.
\end{abstract}
\maketitle
	
\section{INTRODUCTION\label{sec:intro}}
Accumulation of tritium in the plasma facing materials of a fusion reactor, such as ITER, is of primary concern for the successful long-term operation of the installation. The administrative safety restrictions bound the total amount of tritium collected inside the vacuum chamber of the machine to around 700 g during the whole experimental campaign.\cite{roth2008tritium,roth2009recent} The analysis of data on tritium accumulation in the existing fusion devices and extrapolation to future large-scale machines shows that in worst case scenarios, for the beryllium-tungsten tokamak design, the limit can be achieved after several hundred of discharges,\cite{roth2008tritium} resulting in the reactor shutdown for the tritium removal procedures.\cite{zalovznik2020deuterium}

The large volume of the vacuum chamber in the future fusion reactors (e.g., for ITER it will be 840 m$^3$) will require the machine operation without chamber depressurization for a possibly longer time. For this reason, the development of means for the remote control of the tritium content in the machine first wall are a high-priority task. To address this issue, several methods were proposed, viz. the thermal outgassing and first-wall baking for monitoring the global tritium inventory,\cite{federici2001vessel,de2017efficiency} and the laser techniques for controlling the local gas content in the wall materials.\cite{schweer2009laser,huber2011development,philipps2013development,li2016review,maurya2020review,van2021monitoring} While data on the global tritium content is important to assess the overall performance of the installation, special attention is paid to monitoring the local tritium inventory due to sputtering of the reactor plasma-facing components and the subsequent formation of tritium-metal co-deposits in a number of specific areas within the tokamak vacuum chamber.\cite{roth2008tritium,schmid2015walldyn,de2017efficiency}

The laser techniques for the analysis of the tokamak first wall are based on using the laser radiation to remotely heat the surface of the tokamak tiles and register either the radiation signal from the gas cloud/plasma plume formed near the target surface or the particle flux from the tile surface. During analysis, the irradiated sample can undergo modifications, associated with material evaporation, ablation, cracking, etc. The class of methods, which involves material phase transitions, is represented by Laser-Induced Breakdown Spectroscopy\cite{li2016review,paris2017detection,maurya2020review,van2021monitoring,jogi2022laser}, Laser-Induced Ablation optical\cite{huber2011development,philipps2013development,gierse2014situ,hu2018laser,oelmann2021analyses}/mass\cite{oelmann2018depth,li2019quantitative,oelmann2019depth,lyu2022characterization}-spectroscopy. The methods, which do not lead to surface modifications, are based on the Laser-Induced-Desorption with subsequent mass-spectrometry of released gas\cite{huber2001situ,schweer2007situ,schweer2009laser,zlobinski2011laser,widdowson2021evaluation,lyu2021characterization} or by optical spectroscopy in the presence of surrounding plasma.\cite{huber2011development,zlobinski2011laser,zlobinski2011situ} In all methods, the laser pulse striking the surface of an analyzed sample leads to the target heating, which causes the thermal desorption of particles retained in the material into vacuum.

For fusion applications in the ITER tokamak, the LID diagnostic\cite{zlobinski2011laser,de2017efficiency} was proposed as a candidate method due to its simplicity and non-destructive features (the laser pulses do not damage the tiles and contaminate plasma with heavy impurities). The interpretation of LID data, viz. the total number of particles desorbed from the tile during the analysis and the temporal sweeps of the desorbed particle flows to the mass-spectrometer, involves using numerical modeling for the reconstruction of the tritium desorption dynamics. The diagnostic models are built around the set of desorption equations coupled with the heat equation describing the sample temperature dynamics under the pulsed laser irradiation.\cite{spork2013determination,yu2017deuterium,de2017efficiency,gasparyan2021laser,kulagin2022numerical,kulagin2023effect} 

The major assumption put into the desorption codes for modeling LID data is the application of one-dimensional approximation for treating both the particle diffusion and heat conduction in the sample.\cite{zlobinski2011laser,yu2017deuterium,gasparyan2021laser} The one-dimensional theory of heat and particle transport in the solid sample was developed in detail for laser pulses with different shapes and intensities (e.g., see Refs. \onlinecite{jaeger1959conduction,sparks1976theory,sands2011pulsed,prokhorov2018laser} and the references therein) and applied to interpreting diagnostic signals for a wide range of laser-based methods.\cite{kulagin2023effect}

At the same time, in contrast to the conventional thermal desorption experiments, where analyzed samples are heated as a whole, the laser irradiation of a sample is always local. The laser spot is localized in the small area on the sample surface, which, depending on the material thermal properties and the duration of the laser pulse, can potentially lead to deviations from the one-dimensional picture of the heat propagation and, consequently, particle desorption from the sample.

The heat transfer in solids in two- and three-dimensional approximations were considered both analytically\cite{rykalin1957die,kim1990hyperbolic,prokhorov2018laser,chen2017analytical,xu2018theoretical,turkyilmazoglu2021analytic} and numerically\cite{chen2002axisymmetric,toyserkani20043,shuja2007laser,perry2009effect,starikov2015atomistic,kumar2016finite,xiao2018thermal} in connection to technological laser applications. However, there is presently no analysis of the sample temperature dynamics during LID. In particular, there is no assessment of the degree, as to which the one-dimensional heat transport can be applied to modeling the diagnostic signals. In this study, we derive a number of expressions (the rigorous and simplified ones) for the two-dimensional temperature profile established in the sample during LID and assess the validity of using the one-dimensional approximation to describe the sample heating and particle desorption induced by short, millisecond laser pulses characteristic for the diagnostic. In Sec. \ref{sec:preliminaries}, the qualitative estimates for the parameters of the heat transport in the sample under pulsed laser irradiation are given. Sec. \ref{sec:analytical} provides the analytical theory of heat propagation in the semi-infinite sample in one- and two-dimensional approximations. Temperature dynamics in a test tungsten sample obtained in the one- and two-dimensional approximations are shown and discussed in Sec. \ref{sec:approximations}. The sensitivity analysis of the obtained results to experimental uncertainties in the parameters of the laser light and analyzed sample is provided in Sec. \ref{sec:sensitivity_analysis}. The derived expressions for the temperature distributions in the sample are used to assess the importance of taking dimensional effects into account for modeling desorption dynamics of tritium in Sec. \ref{sec:gas_desorption}. The conclusions are summarized in Sec. \ref{sec:conclusions}. Mathematical apparatus necessary to resolve a particular integration problem in Sec. \ref{sec:analytical} is provided in Appendix.

\section{PRELIMINARIES\label{sec:preliminaries}}
Before presenting the analytical results of the study, we first qualitatively assess the importance of different physical processes, involved in setting the heat and particle dynamics in the sample during LID.

\begin{figure}[H]
	\centering
	\includegraphics[scale=0.75]{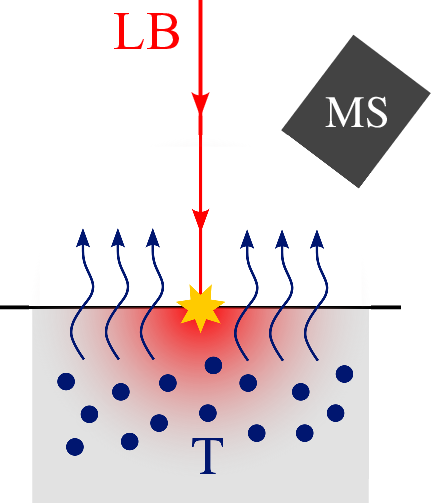}
	\caption{The schematic of the LID diagnostic. The laser beam (LB) heats the sample surface, causing the release of tritium (T) from the defects. Tritium diffuses to the sample surface, desorbs and is collected by the mass-spectrometer (MS).\label{fig:LID}}
\end{figure}

The LID analysis of a target involves irradiating the sample surface with short laser pulses. The target irradiation mobilizes gas particles trapped in the defects of the material structure, which are then released from the sample and collected by the mass-spectrometer (the secondary light emission of desorbed particles can be also measured). Data on the total amount of particles desorbed from the sample and on the temporal dependence of the particle flux are used to reconstruct the parameters of the gas trapping in the sample. The local concentration of gas in the near-surface layers of a sample is of primary concern, since it determines the total amount of tritium accumulated in the fusion reactor. The schematic of the diagnostic is shown in Fig. \ref{fig:LID}.

Dynamics of the particle diffusion, trapping in the material and the subsequent release from the defects and from the sample surface are governed by the target temperature $T$. The characteristic rates of the processes are linked to $T$ through the Arrhenius-like law, $S \propto \exp(-E/kT)$ ($k$ is the Boltzmann constant) and are thus strongly dependent on the local temperature of the sample. For large values of the activation energy $E$, even moderate variations in $T$ can lead to large changes in the particle desorption dynamics, resulting in enhanced, or on the contrary, suppressed release of particles from the analyzed sample.

In LID experiments, lasers with the characteristic intensity $I \sim 500-1000$ MW/m$^2$, laser spot area $S \sim$ several mm$^2$, and the characteristic pulse duration $t_p \sim 1$ ms are used. The total energy in a pulse, $E$, is rather small $E = I S t_p  \lesssim$ several Joules. The samples are heated to high temperatures $\sim 2000-3000$ K, nevertheless not exceeding the material melting temperature. The sample surface is not damaged by the pulses and remains solid during the analysis.

To describe the temperature of the samples during LID experiments, one-dimensional models\cite{jaeger1959conduction} of heat transfer are extensively used. Within these models, the heat is considered propagating along a single direction, into the material bulk, with no heat outflow in the sideways direction. However, the way the energy absorbed by the target is redistributed across the sample must depend on the laser spot area, time of observation, etc, so that the one-dimensional approximation must be valid only in a certain range of physical parameters governing the irradiation process.

The dynamics of the sample temperature are governed by a group of processes, which include the radiation absorption, heat propagation into the material bulk and to the sample surface, and, potentially, to the phase transitions of the sample material. While the first process is responsible for the heating of the sample material, the other ones lead to the sample cooling. In order to construct the model of heat transfer in the target, we assess the importance of each channel of energy redistribution in the material.

In LID experiments, the power of the laser is chosen so as to not increase the sample temperature above the melting point and thus prevent the material from melting, evaporation, ablation, etc. For this reason, below we shall assume that no phase transformations take place in the material. As a result, the cooling of the sample results from the energy flow into the material bulk and to the sample surface, where it is re-radiated backwards into the surrounding medium.

To assess the effectiveness of the sample cooling by the surface radiation, we consider the Stefan-Boltzmann law, $q_s = \varepsilon \sigma T_s^4$, where $q_s$ is the surface radiation flux, $\varepsilon$ is the blackness coefficient, $\sigma$ is the Stefan-Boltzmann constant and $T_s$ is the surface temperature. For the typical LID laser pulse intensity $I \lesssim 1000$ MW/m$^2$, we find that the surface black-body radiation can fully compensate the input laser power, $q_s = I$, if the surface temperature satisfies the condition $T_s \lesssim 10$ kK, which is well beyond the melting point for any of the existing materials. For tungsten, which is considered as the main material for the first wall of a fusion reactor, the ratio $q_s/I$, taken at the melting point $T_s = 3715$ К in the black-body limit $\varepsilon = 1$, yields $q_s/I \sim 10^{-2} \ll 1$, implying that the black-body radiation from the surface can be neglected in the description of heat transfer under the considered conditions. Hence, the main mechanism for the sample cooling is the heat propagation into the bulk of the sample. For this process, we can assume that the heating and cooling of the sample has ended, if the final sample temperature is close to the bulk temperature (i.e. that of the thermostat). To estimate the depth of the heated layer, $l_{hl}$, we can equate the total energy in the pulse to the total thermal energy accumulated in the heated layer, $E = \rho l_{hl} \pi r_0^2 C_p T_0$, where $r_0$ is the laser spot radius, $\rho$ is the material mass density, $C_p$ is the specific heat capacity and $T_0$ is the thermostat temperature. For $E = 4$ J, $r_0 \sim 1$ mm, $T_0 \sim 300$ K, $C_p = 144$ J/(kg$\cdot$K), $\rho = 19\times10^{3}$ kg/m$^3$,\cite{zinoviev1989,zlobinski2011laser} this estimate leads to $l_{hl} \sim 1.6$ mm. At the same time, the estimate of the heat diffusion length, $l_h = 2a\sqrt{t_p}$ [$a^2 = \varkappa/(\rho C_p)$ is the thermal diffusivity, and $\varkappa$ is the thermal conduction coefficient], for a millisecond laser pulse $t_p \sim 1$ ms, irradiating tungsten, $\varkappa = 118$ W/(m$\cdot$K), $a^2 \approx 4.3\times10^{-5}$ m$^2$/s, yields $l_h \sim 0.4$ mm. As seen, all three length scales, $l_{hl}$, $l_h$, and $r_0$ are of the same order of magnitude, $l_{hl} \sim l_h \sim r_0$, implying that the heat propagation in the material can no longer be treated in the one-dimensional approximation. For timescales $t \gtrsim t_p$, the heat not only propagates into the material bulk but spreads across the sample. As a result, it can be envisaged that the temperature profile set in the target will deviate from the one-dimensional approximation at sufficiently large times.

The estimates show that the heat propagates on distance on the order of millimeters into the sample. The particle diffusion dynamics, on the contrary, has a significantly different linear scale. Indeed, the diffusion length is set by $l_d = \sqrt{4Dt}$, where $D$ is the diffusion coefficient. Estimating $D \sim 10^{-7}$ m$^2$/s, for $t \sim t_p \sim 1$ ms we find $l_d \approx 20\ \mu\mathrm{m}$. Thus, $l_h/l_d \sim 20$, and the particle/heat transport occur on significantly disparate spatial scales.

Another important aspect of particle transport in the sample under laser irradiation is the sensitivity of the particle release from the defects to variations in the sample temperature. As already mentioned, the particle de-trapping rate follows the Arrhenius law, $R_{dt} \propto \exp\left(-E_{dt}/kT\right)$, where $E_{dt}$ is the de-trapping energy. Hence, the error in the magnitude of $R$ related to the error in the temperature $T$ can be estimated as $\Delta{R}/R = (E_{dt}/T)(\Delta{T}/T) \equiv (E_{dt}/T)\delta{T}$ ($\delta{\ldots}$ denotes the relative variation). For traps having $E_{dt} \sim 2$ eV and the sample temperature $T \sim 1000-3000$ K (at this temperature, the gas release from the traps becomes effective), we have $\delta{R} \sim (8-23)\delta{T}$, i.e. even a small error in $T$ can produce a substantial error in the detrapping rates. Thus, to have a reasonable accuracy for $R$, e.g. $\delta{R} \leq 0.25$, we must require that $\delta{T} \leq 0.01-0.03$, i.e. the temperature field must be reconstructed with the accuracy no worse than several percent. For traps with lower energies, this restriction becomes relaxed, however remains rather stringent. For $E_{dt} \sim 1$ eV, one has $\delta{R} \leq 0.25$, when $\delta{T} \leq 0.02-0.06$, which is also below the 10\% threshold.

The final comment is related to the form of the heat equation. In the general form, it is written as
\begin{align}
	\rho C_p \ddtp{T} = \nabla\left(\varkappa\nabla{T}\right) + Q. \label{eqn:temp_eqn_nonlinear}
\end{align}
Here, the material density $\rho(T)$, the specific heat $C_p(T)$, and the thermal conductivity coefficient $\varkappa(T)$ depend on the temperature, so that the heat equation is, strictly speaking, non-linear. Despite that, it can be cast into a more conventional form resembling the linear heat equation. Indeed, by applying the Kirchhoff transform and introducing the thermal potential $\theta = \int\varkappa\dif{T} = \theta(T)$, we re-cast Eq. (\ref{eqn:temp_eqn_nonlinear}) as
\begin{align}
	\rho C_p \frac{dg}{d\theta}\ddtp{\theta} = \nabla^2{\theta} + Q.
\end{align}
In this equation $g(\theta) \equiv \theta^{-1}$, $\rho = \rho(\theta)$, and $C_p=C_p(\theta)$. By denoting $a_\theta^2(\theta) = 1/(\rho C_p dg/d\theta)$ as the effective diffusivity for the thermal potential and $Q' = Q/(\rho C_p dg/d\theta)$, we straightforwardly re-cast the equation to the linear-like form
\begin{align}
	\ddtp{\theta} = a_\theta^2\nabla^2{\theta} + Q'. \label{eqn:theta_eqn}
\end{align}

\begin{figure}
	\includegraphics[scale=0.75]{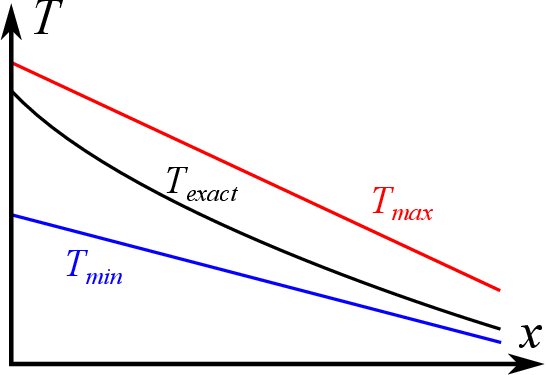}
	\caption{The schematic representation of the temperature profile in case of the material thermal properties dependent on $T$. The bounding curves represent the temperature profiles obtained with the thermal properties taken at some constant limiting temperatures $T_{\max}$ and $T_{\min}$.\label{fig:temp_profile_qualitative}}
\end{figure}

The derived equation has an important implication. Since it has the form similar to the linear heat equation, in which the thermal properties, $\varkappa$, $C_p$, $\rho$, $a^2 = \varkappa/(\rho C_p)$ are independent of $T$,
\begin{align}
	\ddtp{T} = a^2 \nabla^2{T} + \frac{Q}{\rho C_p}, \label{eqn:T_eqn}
\end{align}
solutions to Eqs. (\ref{eqn:theta_eqn}) and (\ref{eqn:T_eqn}) must be similar. Herein, since in Eq. (\ref{eqn:theta_eqn}) the effective thermal properties, $a_\theta$, $C_p$, $\rho$ vary with $\theta$, the solution to Eq. (\ref{eqn:theta_eqn}) must be bounded by the solutions to Eq. (\ref{eqn:T_eqn}), obtained for the values of $\varkappa$, $C_p$, $\rho$ taken at some limiting values $T_{\max}$ and $T_{\min}$, as schematically shown in Figure \ref{fig:temp_profile_qualitative}.

Thus to find the analytical solution for the temperature distribution in the sample, we do not need to solve the exact non-linear equation (\ref{eqn:theta_eqn}). Instead, we can solve Eq. (\ref{eqn:T_eqn}) and then analyze the solution by varying the thermal properties in a chosen temperature range.

\section{ANALYTICAL FORM OF TEMPERATURE DISTRIBUTION IN TARGET\label{sec:analytical}}
To analyze the heat propagation in the sample, we shall use the following linear heat equation for the sample temperature $T$:
\begin{align}
	\rho C_p \ddtp{T} = \varkappa \nabla^2{T} + Q. \label{eqn:heat_pnitial}
\end{align}
Here $\rho$ is the material mass density, $C_p$ is the specific heat, and $\varkappa$ is the thermal conductivity of the sample material. In what follows, we shall keep these parameters constant, assuming that they are independent of the sample temperature $T$. The source $Q$ is the heating source due to the laser irradiation of the sample. The distribution of $Q$ is governed by the Beer-Bouguer-Lambert law, 
\begin{align}
	Q = I \alpha \exp(-\alpha s) Y(t). \label{expr:Q}
\end{align}

The laser intensity $I$ is the effective intensity of the laser pulse absorbed by the target, taking into the account the reflection of light from the optical elements and the sample surface, i.e. $I = I_0 (1-R)$, where $I_0$ is the initial laser pulse intensity, and $R$ is the collective reflection coefficient for the experimental setup. Since $I_0$ and $R$ can be absorbed into $I$ (they are not used in any other relation other than for $I$), in what follows we shall keep the above notation for both $I$ and $Q$. For the spatial distribution of $I$, we shall use the Gaussian profile
\begin{align}
	I(r) = I_m \exp\left(-\frac{r^2}{r_0^2}\right),
\end{align}
where $I_m$ is the maximum intensity in the center of the laser spot, $r_0$ is the effective radius of the spot, and $r$ is the distance from the spot center to the observation point. The effective spot cross-section is given by the relation $S = 2\pi\int_0^{+\infty}[I(r)/I_m]r\dif{r} = \pi r_0^2$.

The parameter $\alpha$ is the extinction coefficient, defined as
\begin{align}
	\alpha = \frac{4\pi k}{\lambda},
\end{align}
where $k$ is the factor associated with the radiation absorption, and $\lambda$ is the wavelength of the laser light. For the infra-red laser used for LID experiments, having $\lambda = 1064$ nm,\cite{zlobinski2011laser} and tungsten as an analyzed sample, one has $k \approx 11$.\cite{ordal1988optical} The radiation attenuation length is $1/\alpha \approx 8$ nm, which is much smaller than the scale $l_h$ (and also the diffusion length scale $l_d$). Hence, we can make the passage to the limit $1/\alpha \rightarrow 0$, and obtain $\lim_{1/\alpha \rightarrow 0}\alpha \exp(-\alpha s) = 2\delta(s)$, where $\delta(s)$ is the Dirac delta-function (the details of the limit transition are covered in Appendix). The parameter $s$ is the distance measured along the ray, from the surface into the bulk of the sample.

The function $Y(t)$ describes the temporal profile of the laser pulse. For LID experiments, it can be assumed to have a trapezoid form,\cite{zlobinski2013hydrogen,zlobinski2019laser,kulagin2022numerical,kulagin2023effect} i.e.
\begin{align}
	Y(t) = H(t_p - t)\left(1 - \Delta \frac{t}{t_p}\right), \label{expr:time_function}
\end{align}
where $0 < \Delta < 1$ is the coefficient describing the attenuation of the laser pulse, and $H$ is the Heaviside function. 

\begin{figure}[H]
	\centering
	\includegraphics[scale=0.9]{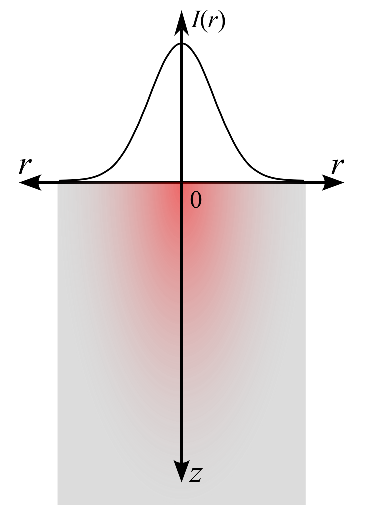}
	\caption{The sketch of the problem geometry and reference system used for the analysis of sample temperature dynamics.\label{fig:sketch}}
\end{figure}

We shall use the cylindrical reference system. The reference point will be located at the laser spot center on the sample surface. The symmetry axis $z$ will be directed normally to the sample surface, which we shall consider flat. The distance $r$ and the angle $\varphi$ are measured regularly, in the azimuthal plane perpendicular to the axis $z$. In this reference frame, the sample will be treated as semi-infinite, bounded by the plane $z = 0$. It will be assumed that the laser light comes to the sample at the normal incidence, so that the distance $s$ in the relation (\ref{expr:Q}) is equal to the coordinate $z$. The sketch of the reference system and problem geometry are shown in Fig. \ref{fig:sketch}.

With the above considerations, Eq. (\ref{eqn:heat_pnitial}) can be written as follows:
\begin{align}
	\ddtp{T} = a^2 \left(\frac{\partial^{2}T}{\partial{r^2}} + \frac{1}{r}\frac{\partial{T}}{\partial{r}} + \frac{\partial^{2}T}{\partial{z^2}}\right) + P, \label{eqn:heat_full}
\end{align}
where $a^2 = \varkappa/(\rho C_p)$ is the thermal diffusivity, 
\begin{align}
	P(r,z,t) = P_0 \exp\left(-\frac{r^2}{r_0^2}\right)  \delta(z) H(t_p - t)\left[1 - \Delta \frac{t}{t_p}\right],
\end{align}
$P_0 = 2I_m / (\rho C_p)$, and we have additionally taken into account the axial symmetry of the temperature profile $T = T(r,z,t)$.

The boundary conditions imposed on the temperature profile are:
\begin{gather}
	\lim_{r\rightarrow \infty} T(r,z,t) = \lim_{z\rightarrow \infty} T(r,z,t) = T_0,\\
	\frac{\partial{T}}{\partial{z}}(r,0,t) = 0.
\end{gather}
The first group of equations describes the thermostat kept at the constant temperature $T_0$. The second equation represents the closed boundary, allowing no radiation or other energy sinks at the sample surface. The condition can be imposed not only on the sample surface but also at the thermostat boundary, i.e. $\nabla{T}|_{r,z\rightarrow \infty} = 0$, due to the energy conservation condition $\nabla\cdot\mathbf{q} = 0$, implying, in a non-one-dimensional case, the diminishing of the heat flux $\mathbf{q} = -\varkappa\nabla{T}$ with the distance from the laser irradiated spot on the sample surface.

The initial condition is given by
\begin{align}
	T(r,z,0) = T_0.
\end{align}

To solve Eq. (\ref{eqn:heat_full}), we introduce the auxiliary variable $T^*(r,z,t)$:
\begin{align}
	T^*(r,z,t) = T(r,z,t) - T_0.
\end{align}
The heat equation, boundary and initial conditions are re-cast as
\begin{gather}
	\ddtp{T^*} = a^2 \left(\frac{\partial^{2}T^*}{\partial{r^2}} + \frac{1}{r}\frac{\partial{T^*}}{\partial{r}} + \frac{\partial^{2}T^*}{\partial{z^2}}\right) + P, \label{eqn:heat_full_reduced}\\
	\lim_{r\rightarrow \infty} T^*(r,z,t) = \lim_{z\rightarrow \infty} T^*(r,z,t) = 0,\\
	\frac{\partial{T^*}}{\partial{z}}(r,0,t) = 0,\\
	T^*(r,z,0) = 0.
\end{gather}

The Green's function for the transformed problem is given by the relation (in its writing, the Cartesian representation was used)\cite{sveshnikov2004lectures}
\begin{align}\label{expr:Green_function}
\begin{split}
	&G(x,y,z,\xi,\eta,\zeta,t-\tau) = \frac{1}{[2a\sqrt{\pi(t-\tau)}]^{3}}\\
	&\qquad\times
	\begin{cases}
		\exp\left[-\frac{(x-\xi)^2 + (y-\eta)^2 + (z - \zeta)^2}{4a^2(t-\tau)}\right] + \exp\left[-\frac{(x-\xi)^2 + (y-\eta)^2 + (z + \zeta)^2}{4a^2(t-\tau)}\right], & t - \tau > 0, \\
		0, & t - \tau \leq 0,
	\end{cases}
\end{split}
\end{align}
where $x$ and $y$ are the mutually perpendicular axes in the azimuthal plane.

By using the Green's function, the expression for the temperature profile in the integral form can be written as\cite{sveshnikov2004lectures} (in writing of the relation the backwards transition from $T^*$ to $T$ was made)
\begin{align}
	\begin{aligned}
	T(r,z,t) &= T_0 + \int_{0}^{t} \dif{\tau} \iint_{-\infty}^{+\infty} \dif{\xi}\dif{\eta}\\
	&\int_{0}^{+\infty}\dif\zeta G(x,y,z,\xi,\eta,\zeta,t-\tau)P(\xi,\eta,\zeta,\tau).
	\end{aligned}
\end{align}

The integration over $\zeta$ yields
\begin{align}
\begin{aligned}
	T(r,z,t) = T_0 + P_0&\int_{0}^{t} \frac{Y(\tau)}{[2a\sqrt{\pi(t-\tau)}]^{3}} \dif{\tau}\\
	&\iint_{-\infty}^{+\infty}\dif{\xi}\dif{\eta}\exp\left[-\frac{(x-\xi)^2 + (y-\eta)^2 + z^2}{4a^2(t-\tau)}\right]\exp\left(-\frac{\xi^2 + \eta^2}{r_0^2}\right). \label{expr:temp_intermediate}
\end{aligned}
\end{align}

To simplify the obtained relation, we consider an auxiliary integral
\begin{align}
	I_x = \int_{-\infty}^{+\infty}\exp\left[-\frac{(x-\xi)^2}{4a^2(t-\tau)} - \frac{\xi^2}{r_0^2}\right]\dif{\xi}.
\end{align}
By introducing the parameters $L_h = [4a^2(t-\tau)]^{1/2}$, $1/p^2 = 1/L_h^2 + 1/r_0^2$, the exponent argument can be represented as
\begin{align}
	\frac{(x-\xi)^2}{L_h^2} + \frac{\xi^2}{r_0^2} = h_x + \left(\frac{px}{L_h^2} - \frac{\xi}{p}\right)^2,
\end{align}
where 
\begin{align}
	h_x = \left(\frac{L_h^2}{p^2} - 1\right)\frac{(px)^2}{L_h^4}.
\end{align}
As a result, we have
\begin{align}
	I_x = p e^{-h_x} \int_{-\infty}^{+\infty}e^{-\omega_x^2}\dif{\omega_x} = \sqrt{\pi}pe^{-h_x},
\end{align}
where $\omega_x = \xi/p - px/L_h^2$.

Analogous calculations yield
\begin{align}
	\begin{aligned}
	I_y = \int_{-\infty}^{+\infty}\exp\left[-\frac{(y-\eta)^2}{4a^2(t-\tau)} - \frac{\eta^2}{r_0^2}\right]\dif{\eta} = p e^{-h_y} \int_{-\infty}^{+\infty}e^{-\alpha_y^2}\dif{\alpha_y} = \sqrt{\pi}pe^{-h_y},
	\end{aligned}
\end{align}
where
\begin{align}
	h_y = \left(\frac{L_h^2}{p^2} - 1\right)\frac{(py)^2}{L_h^4}.
\end{align}
By substituting the relations for $I_x$ and $I_y$ into Eq. (\ref{expr:temp_intermediate}), after some algebra we arrive at
\begin{align}
	T(r,z,t) = T_0 + \pi P_0\int_{0}^{t}\dif{\tau} \frac{Y(\tau)}{[2a\sqrt{\pi(t-\tau)}]^{3}} p^2 \exp\left[-\frac{z^2}{4a^2(t-\tau)} - h_x - h_y\right].
\end{align}
By expanding
\begin{gather}
	p^2 = \left(\frac{1}{L_h^2} + \frac{1}{r_0^2}\right)^{-1} = \frac{r_0^2}{1 + \frac{r_0^2}{4a^2(t-\tau)}},\\
	\begin{aligned}
		h_x + h_y &= \left(\frac{L_h^2}{p^2} - 1\right)\frac{p^2}{L_h^4}(x^2 + y^2) = \frac{L_h^2 - p^2}{L_h^4}(x^2 + y^2) \\
		&= \left(1 - \frac{r_0^2}{L_h^2 + r_0^2}\right)\frac{(x^2 + y^2)}{L_h^2} = \frac{x^2 + y^2}{4a^2(t-\tau) + r_0^2},
	\end{aligned}
\end{gather}
recalling that $r^2 = x^2 + y^2$, $P_0 = 2I_m/(\rho C_p)$, and $a = \sqrt{\varkappa/(\rho C_p)}$, we finally obtain
\begin{align}
	T(r,z,t) = T_0 + \frac{I_m r_0^2}{\sqrt{\pi \rho C_p \varkappa}} \int_{0}^{t} \dif{\tau} \frac{Y(\tau)}{\sqrt{t - \tau}} \frac{\exp\left[-\frac{z^2}{4a^2(t-\tau)} - \frac{r^2}{4a^2(t-\tau) + r_0^2}\right]}{r_0^2 + 4a^2(t-\tau)}.\label{expr:T_final}
\end{align}
The derived relation defines the spatial and temporal dependence of the temperature field $T$. It coincides with the general expression (2.35) given in Ref. \onlinecite{prokhorov2018laser}. In what follows, we shall consider the trapezoid pulse shape profiles and deduce a number of new analytical expressions for the temperature distribution in the sample, other than those covered in Ref. \onlinecite{prokhorov2018laser}. From Ref. \onlinecite{rykalin1957die}, the derived expression differs in that it treats the heat transport via the Green's function taking into account the deviation of the thermal front from the spherical shape during the propagation of heat. In comparison to studies that employ the Laplace transform to solve the heat equation,\cite{kim1990hyperbolic,chen2017analytical,xu2018theoretical} our result differs in that the temperature profile (\ref{expr:T_final}) is given by the Green's, rather than the inverse Laplace integral distribution. As will be shown below, the derived expressions can be further reduced to a simplified form expressed in the terms of elementary functions. In addition, we shall apply formula (\ref{expr:T_final}) to analysis of heat propagation driven by the trapezoid laser pulse, whereas the studies\cite{chen2017analytical,xu2018theoretical} consider the sample heating by a train of instantaneous laser pulses.  In what follows, Eq. (\ref{expr:T_final}) will be used as the starting point for analysis of the sample temperature dynamics during LID.

We now recover the well-known expression for the surface temperature in the one-dimensional limit.\cite{jaeger1959conduction} By setting $z = 0$ and making the transition $r \rightarrow \infty$, we deduce
\begin{align}
	T(r,z,t) = T_0 + \frac{I_m}{\sqrt{\pi \rho C_p \varkappa}}\int_{0}^{t}\dif{\tau} \frac{Y(\tau)}{\sqrt{t - \tau}}.\label{expr:T_final_1d}
\end{align}
From physical standpoint, the one-dimensional approximation is valid if $r_0 \gg 2a\sqrt{t - \tau}$ or, equivalently, when $t - \tau \ll [r_0/(2a)]^2$. By estimating $r_0 \sim 1$ mm, for tungsten [$\varkappa = 118$~W/(m$\cdot$K), $\rho = 19.1\times10^3$~kg/m$^3$, $C_p = 144.5$~J/(kg$\cdot$K)] one has $t - \tau \ll 6\ \mathrm{ms} \sim t_p$. Hence, the one-dimensional approximation is valid only during the initial stages of the target heating. At larger timescales (e.g. on the order of the LID laser pulse duration), the one-dimensional approximation breaks and the heat transfer and temperature dynamics of the target must be treated in the two-dimensional approximation.

We re-cast Eq. (\ref{expr:T_final}) to the form, suitable for the further analysis. By introducing new variables $\theta = 2a\sqrt{t - \tau}/r_0$, $\rho = r/r_0$, and $\lambda = z/L$ ($L$ is the normalization parameter that we choose as $L \sim l_d \sim 10$ $\mu$m), substituting $Y(\tau)$ with (\ref{expr:time_function}), the expression for $T(r,z,t)$ is re-cast in the dimensionless form as
\begin{align}
	\frac{T(r,z,t)}{T_0} = 1 + J_0 e^{-\rho^2} \int_{\theta_{\min}}^{\theta_{\max}} \dif{\theta} \left(1 - \Delta\frac{\theta_{\max}^2 - \theta^2}{\theta_p^2}\right)\frac{\exp\left(-\frac{L^2}{r_0^2}\frac{\lambda^2}{\theta^2} + \frac{\theta^2}{1 + \theta^2}\rho^2\right)}{1 + \theta^2}, \label{expr:T_final_norm}
\end{align}
where $J_0 = (I_m r_0)/(T_0\varkappa\sqrt{\pi})$, $\theta_{\max} = 2a\sqrt{t}/r_0$, and $\theta_p = 2a\sqrt{t_p}/r_0$. The parameter $\theta_{\min} = 0$ if $t \leq t_p$, otherwise $\theta_{\min} = 2a\sqrt{t - t_p}/r_0$. The obtained relation is, so far, the exact form of the temperature profile. In its deriving, no further assumptions on the pulse temporal shape were made, so it can be equally applied to millisecond or shorter (microsecond, etc.) laser pulses, provided i) the pulse profile is trapezoid, ii) the surface temperature dynamics remains the same (no phase transition, black-body radiation, etc).

As a next step, we assess the importance of retaining the dependence of $T$ on the normalized depth $\lambda$. From Eq. (\ref{expr:T_final_norm}), it follows that the $\lambda$ contribution is important if $(L/r_0)^2\lambda^2/\theta^2 \gtrsim \rho^2\theta^2/(1+\theta^2)$, or when $\lambda \gtrsim [\theta^2/(1+\theta^2)^{1/2}](r_0/L)\rho$. Estimating $L \sim 10$ $\mu$m, $r_0 \sim 1$ mm, $t - \tau \sim 1$ ms, and using the tungsten thermal properties, one has $\lambda \gtrsim 16\rho$. At larger times, $t - \tau \sim 10$ ms, the inequality becomes $\lambda \gtrsim 10^2\rho$. Thus the depth dependence in Eq. (\ref{expr:T_final_norm}) can be dropped out. This results explains why the dynamics of gas desorption from the near-surface layer is almost insensitive to the details of the heat transfer in the bulk of the sample and is mainly determined by the target surface temperature.

Now we transform the integrand term under the parentheses to a more compact form, by replacing
\begin{align}
	1 - \Delta\frac{\theta_{\max}^2 - \theta^2}{\theta_p^2} = K_1 + K_2 \theta^2,
\end{align}
where $K_1 = 1 - \Delta(\theta_{\max}/\theta_p)^2$, $K_2 = \Delta/\theta_p^2$. As a result, Eq. (\ref{expr:T_final_norm}) can be transformed to
\begin{align}
	\frac{T(r,z,t)}{T_0} = 1 + J_0 e^{-\rho^2} \int_{\theta_{\min}}^{\theta_{\max}} \dif{\theta} \left(K_1 + K_2 \theta^2\right)\frac{\exp\left(\frac{\theta^2}{1 + \theta^2}\rho^2\right)}{1 + \theta^2}, \label{expr:T_final_norm_reduced}
\end{align}

To further simplify the integral relation, we introduce the auxiliary function
\begin{align}
	F(\rho,\theta) = \int_{0}^{\theta} \dif{\varphi} \left(K_1 + K_2 \varphi^2\right)\frac{\exp\left(\frac{\varphi^2}{1 + \varphi^2}\rho^2\right)}{1 + \varphi^2}. \label{expr:F}
\end{align}
With its aid, Eq. (\ref{expr:T_final_norm_reduced}) is cast as
\begin{align}
	\frac{T(r,z,t)}{T_0} = 1 + J_0 e^{-\rho^2} \left[F(\rho, \theta_{\max}) - F(\rho, \theta_{\min})\right], \label{expr:T_final_norm_reduced_F}
\end{align}
The function $F$ cannot be resolved as a combination of the elementary or special functions, however the power series representation can be obtained. To do that, we expand the exponent in the Taylor series
\begin{align}
	\exp\left(\frac{\theta^2}{1 + \theta^2}\rho^2\right) = \sum_{n = 0}^{\infty}  \frac{\theta^{2n}}{(1 + \theta^2)^n} \frac{\rho^{2n}}{n!}.
\end{align}
By substituting this relation into (\ref{expr:F}), after some algebra one finds
\begin{align}
	F(\rho,\theta) = \sum_{n = 0}^{\infty}\rho^{2n}\left[K_1 A_n(\theta) + K_2 B_n(\theta)\right], \label{expr:F_series}
\end{align}
where
\begin{align}
	A_n(\theta) = \frac{1}{n!} \int_{0}^{\theta} \dif{\varphi} \frac{\varphi^{2n}}{(1 + \varphi^2)^{n+1}},\\
	B_n(\theta) = \frac{1}{n!} \int_{0}^{\theta} \dif{\varphi} \frac{\varphi^{2(n+1)}}{(1 + \varphi^2)^{n+1}}.
\end{align}
The first several polynomial coefficients $A_n$, $B_n$ ($n = \overline{0-5}$) are given below
\begin{align}
	A_0(\theta) &= \arctan\theta,\\
	A_1(\theta) &= \frac12\arctan(\theta) - \frac{\theta}{2(\theta^2+1)},\\
	A_2(\theta) &= \frac{3}{16}\arctan\theta - \frac{\theta(5\theta^2 + 3)}{16(\theta^2+1)^2},\\
	A_3(\theta) &= \frac{5}{96}\arctan\theta - \frac{\theta(33\theta^4 + 40\theta^2 + 15)}{288(\theta^2+1)^3},\\
	A_4(\theta) &= \frac{35}{3072}\arctan\theta - \frac{\theta(279\theta^6 + 511\theta^4 + 385\theta^2 + 105)}{9216(\theta^2+1)^4},\\
	A_5(\theta) &= \frac{63}{30720}\arctan\theta - \frac{\theta(965\theta^8 + 2370\theta^6 + 2688\theta^4 + 1470\theta^2 + 315)}{153600(\theta^2+1)^5},\\
	B_0(\theta) &= \theta - \arctan\theta,\\
	B_1(\theta) &= \theta + \frac{\theta}{2(\theta^2+1)} - \frac32\arctan\theta,\\
	B_2(\theta) &= \frac12\theta + \frac{\theta(9\theta^2 + 7)}{16(\theta^2+1)^2} - \frac{15}{16}\arctan\theta,\\
	B_3(\theta) &= \frac16\theta + \frac{\theta(87\theta^4 + 136\theta^2 + 57)}{288(\theta^2+1)^3} - \frac{35}{96}\arctan\theta.\\
	B_4(\theta) &= \frac{1}{24}\theta + \frac{\theta(325\theta^6 + 765\theta^4 + 643\theta^2 + 187)}{3072(\theta^2+1)^4} - \frac{315}{3072}\arctan\theta.\\
	B_5(\theta) &= \frac{1}{120}\theta + \frac{\theta(4215\theta^8 + 13270\theta^6 + 16768\theta^4 + 9770\theta^2 + 2185)}{153600(\theta^2+1)^5} - \frac{693}{30720}\arctan\theta.
\end{align}

In the opposite, one-dimensional approximation, the general formula (\ref{expr:T_final_norm}) is re-cast as
\begin{align}
	\frac{T(r,z,t)}{T_0} = 1 + J_0 e^{-\rho^2} \left[G(\lambda,\theta_{\max}) - G(\lambda, \theta_{\min})\right], \label{expr:T_final_1d_norm}
\end{align}
where
\begin{align}
	G(\lambda, \theta) = \int_{0}^{\theta} \dif{\varphi} \left(K_1 + K_2 \varphi^2 \right)\exp\left(-\frac{L^2}{r_0^2}\frac{\lambda^2}{\varphi^2}\right).
\end{align}
Evaluation of the integral yields
\begin{align}
	\begin{aligned}
	G(\lambda, \theta) = K_1 \theta\left[\exp(-s^2) - \sqrt{\pi} s \mathrm{erfc}(s)\right] 
	+\frac{K_2}{3}\theta^3\left[\left(1 - 2s^2\right)\exp(-s^2) - 2\sqrt{\pi}s^3\mathrm{erfc}(s)\right],
	\end{aligned}
\end{align}
where $s = z/(2a\sqrt{t})$, and $\mathrm{erfc}(s)$ is the complementary error function. In the limit $\Delta = 0$, one has $K_1 = 1, K_2 = 0$ and $G$ takes the form
\begin{align}
	G(\lambda, \theta) &= \theta\left[\exp(-s^2) - \sqrt{\pi}s \mathrm{erfc}(s)\right].
\end{align}
For the laser spot center ($\rho=0$) at times $t \leq t_p$, we arrive at the know relation\cite{zlobinski2011laser}
\begin{align}
	T(z,t) = T_0 + 2\frac{I_m\sqrt{t}}{\sqrt{\pi\varkappa\rho C_p}} \left[e^{-s^2} - \sqrt{\pi}s\mathrm{erfc}(s)\right].
\end{align}

It must be emphasized that the relations (\ref{expr:T_final_1d}), (\ref{expr:T_final_1d_norm}) should be termed as quasi-one-dimensional rather than one-dimensional, since they retain the dependence on the radial distance $\rho$. Since the rigorous one-dimensional approximation will not be considered further on, the terms one-dimensional and quasi-one-dimensional will be used interchangeably.

It is also useful to obtain the relation for the maximum temperature $T_{\max}$ of the sample. From the integral form of the temperature profile (\ref{expr:T_final_norm}), it follows that $T_{\max}$ is achieved at the end of the laser pulse. Indeed, for $t \leq t_p$ -- $\theta_{\min} = 0$, $\theta_{\max} \propto \sqrt{t}$ increases with time, and, given that the integrand is positive, the integral grows in time. For $t > t_p$, $\theta_{\min} \xrightarrow{t \rightarrow \infty} \theta_{\max}-0$, so that the integral value must diminish with time (the upper and lower bounds tend to each other). Hence, for the maximum temperature of the sample (which is achieved in the center of the laser spot, $\lambda = \rho = 0$) we find
\begin{align}
	\frac{T_{\max}^{2\mathrm{D}}}{T_0} = 1 + J_0 \int_{0}^{\theta_{\max}(t_p)} \dif{\theta} \left[1 - \Delta\frac{\theta_{\max}^2(t_p) - \theta^2}{\theta_p^2}\right]\frac{1}{1 + \theta^2}. \label{expr:Tmax2D}
\end{align}
By using the power series representation (\ref{expr:T_final_norm_reduced_F}), (\ref{expr:F_series}), this equation is reduced to
\begin{align}
	\frac{T_{\max}^{2\mathrm{D}}}{T_0} = 1 + J_0 \left[K_1 A_0(\theta_{\max}) + K_2 B_0(\theta_{\max})\right]\Big.\Big|_{t=t_p},
\end{align}
or, in the explicit dimensional form,
\begin{align}
	T_{\max}^{2\mathrm{D}} = T_0 + \frac{I_m r_0}{\varkappa\sqrt{\pi}} \Biggl\{(1-\Delta)\arctan\left(\frac{2a\sqrt{t_p}}{r_0}\right) + \Delta\frac{r_0^2}{4a^2t_p} \left[ \frac{2a\sqrt{t_p}}{r_0} - \arctan\left(\frac{2a\sqrt{t_p}}{r_0}\right) \right] \Biggr\}. \label{expr:max_temp}
\end{align}
In the limit of the small spot radius, $r_0/(2a\sqrt{t_p}) = \alpha \ll 1$, by retaining terms $\sim \alpha^0$ and assuming $\Delta < 1$, the expression for the maximal sample surface temperature is reduced to
\begin{align}
	T_{\max}^{2\mathrm{D}} \approx T_0 + (1-\Delta)\sqrt{\frac{\pi}{2}}\frac{I_m r_0}{\varkappa}, \label{expr:max_temp_2d}
\end{align}
i.e., when $I_m = \const$, $T_{\max}$ is linearly dependent on $r_0$ and is independent of the pulse duration $t_p$. If the total energy expenditure is conserved in the pulse, i.e. $I_m r_0^2 = E_m = \const$, then 
\begin{align}
	T_{\max}^{2\mathrm{D}} = T_0 + (1-\Delta)\sqrt{\frac{\pi}{2}}\frac{E_m}{\varkappa r_0}, \label{expr:max_temp_2d_const_Em}
\end{align}
and $T_{\max}$ is inversely proportional to $r_0$.

Similar calculations of $T_{\max}$ in the one-dimensional approximation [i.e. by dropping out the $\theta^2$ term in the denominator under the integral sign in (\ref{expr:Tmax2D})] yield
\begin{align}
	T_{\max}^{1\mathrm{D}} = T_0 + \frac{2}{\sqrt{\pi}}\frac{I_m \sqrt{t_p}}{\sqrt{\varkappa\rho C_p}} \left(1 - \frac23\Delta\right). \label{expr:max_temp_1d}
\end{align}
As seen, in contrast to the two-dimensional case, $T_{\max}^{1\mathrm{D}}$ is independent of $r_0$ and is proportional to $\sqrt{t_p}$, when $I_m = \const$. For $E_m = \const$, one has
\begin{align}
	T_{\max}^{1\mathrm{D}} = T_0 + \frac{2}{\sqrt{\pi}}\frac{E_m \sqrt{t_p}}{r_0^2 \sqrt{\varkappa\rho C_p}} \left(1 - \frac23\Delta\right), \label{expr:max_temp_1d_const_Em}
\end{align}
so that $T_{\max}$ grows inversely proportional to $r_0^2$. 

The observed discrepancies between the two limits are the dimensional effect. In the one-dimensional approximation, the input energy has only one degree of freedom and flows along a single axis. As a result, for the heat flux $\mathbf{q}$ one has $\mathbf{q} = \const$. In the two-dimensional case, the energy is redistributed in space. At large distances from the laser spot, as mentioned in Sec. \ref{sec:preliminaries}, the conservation of the energy flow requires that $q \rightarrow 0$, since as the total surface, through which the energy flows, grows. For $r,z \gg r_0$, the heated area on the sample surface can be treated as a point, and at large distances from the center of the laser spot, $R\rightarrow\infty$, one has $q \propto 1/R^2 \xrightarrow{R\rightarrow\infty} 0$.

\section{QUASI-1D VS. 2D APPROXIMATIONS\label{sec:approximations}}
In the previous section \ref{sec:analytical}, we have obtained the relations for the temperature distribution in the sample, both in quasi-one- and two-dimensional approximations. We now test the derived relations to assess the importance of taking dimensional effects into account and to evaluate the validity of the employed approximations.

For the tests, we employ the laser and sample parameters relevant for the LID experiment.\cite{zlobinski2011laser} The maximum laser intensity $I_m$, the pulse duration $t_p$, the laser spot radius $r_0$, the attenuation parameter for the pulse $\Delta$ are as follows: $I_m = 850$ MW/m$^2$, $t_p = 2.95$ ms,  $r_0 = 1.3$ mm, $\Delta = 0$. For the sample material, tungsten is used. Its thermal properties are:\cite{zlobinski2011laser} $\varkappa = 118$~W/(m$\cdot$K), $\rho = 19.079\times10^3$~kg/m$^3$, $C_p = 144$~J/(kg$\cdot$K) (these values correspond to sample temperature $T = 1000$ K).\cite{zinoviev1989} The depth of the analysis $L$ will be taken corresponding to the diffusion scale $l_d$, i.e. $L = 10$ $\mu$m.

\begin{figure}[H]
	\includegraphics{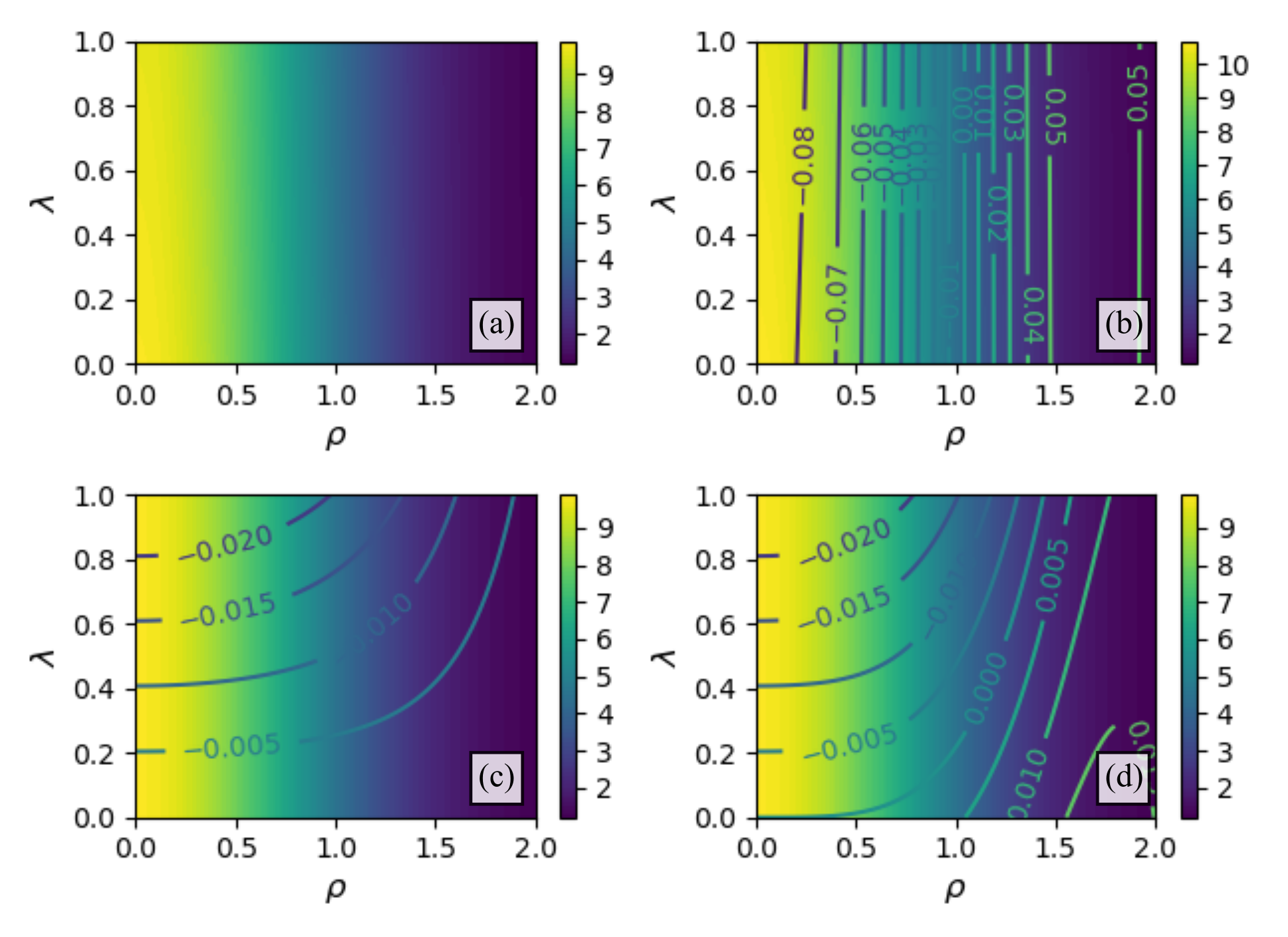}
	\caption{The temperature distribution in the sample obtained at $t = t_p = 2.95$ ms with (\textit{a}) the exact expression (\ref{expr:T_final_norm}), (\textit{b}) the normalized quasi-1D form (\ref{expr:T_final_1d_norm}), (\textit{c}) the $\lambda$-reduced 2D form (\ref{expr:T_final_norm_reduced}), (\textit{d}) the 2D power series approximation (\ref{expr:T_final_norm_reduced_F}). The tungsten properties are $\varkappa = 118$~W/(m$\cdot$K), $\rho = 19.079\times10^3$~kg/m$^3$, $C_p = 144$~J/(kg$\cdot$K). \label{fig:temp_fields_Zlobinskii_case}}
\end{figure}

Figure \ref{fig:temp_fields_Zlobinskii_case} shows the temperature distribution $T/T_0$ in the sample, obtained in the end of the laser pulse, $t = t_p$, by using four different approximations: the exact expression (\ref{expr:T_final_norm}), the normalized 1D form (\ref{expr:T_final_1d_norm}), the $\lambda$-reduced 2D form (\ref{expr:T_final_norm_reduced}) and the 2D power series approximation (\ref{expr:T_final_norm_reduced_F}) obtained by using only four terms, having $n=0,1$, in the expansion for the function $F$. The contour lines superimposed over the heat maps are the isolines showing the relative error $\delta{T} = (T_{ref} - T)/T_{ref}$ of the reconstructed temperature distributions compared to the reference distribution $T_{ref}$ given by (\ref{expr:T_final_norm}).

As seen, the quasi-one-dimensional approximation to the temperature profile shows the largest relative error and overestimates the sample temperature by almost 10\% in the central area of the laser spot, where the temperature takes the largest values. The two-dimensional $\lambda$-reduced form (\ref{expr:T_final_norm_reduced}) shows higher accuracy, overestimating the rigorous temperature profile by no more than 2\%. The power series approximation also shows good accuracy almost identical to the 2D $\lambda$-reduced one, although at large $\rho$ the deviation from the reference distribution increases, indicating the importance of taking higher-order polynomials into account. However, for large values of $\rho$ the sample temperature becomes comparable to that of the thermostat, having, as a consequence, a considerably lesser impact on the gas desorption dynamics from the sample, compared to the hot central spot.

The relative error reaching almost 10\% in the hot central spot of the pulse found in the 1D case, as mentioned in Sec. \ref{sec:preliminaries}, must have a considerable impact on the desorption dynamics, leading to the overestimation of the gas released from the sample having defects with high trapping energies. Note that the isolines showing the relative error distribution propagate up to the sample surface, where the desorption process starts. This would force the gas desorption and enhance the particle loss from the sample near-surface area. This is in contrast, e.g., to the two-dimensional approximation, which shows the increase of $\delta{T}$ with the depth $\lambda$ rather radius $\rho$. More details are given in Sec. \ref{sec:gas_desorption}.

\begin{figure}[H]
	\includegraphics{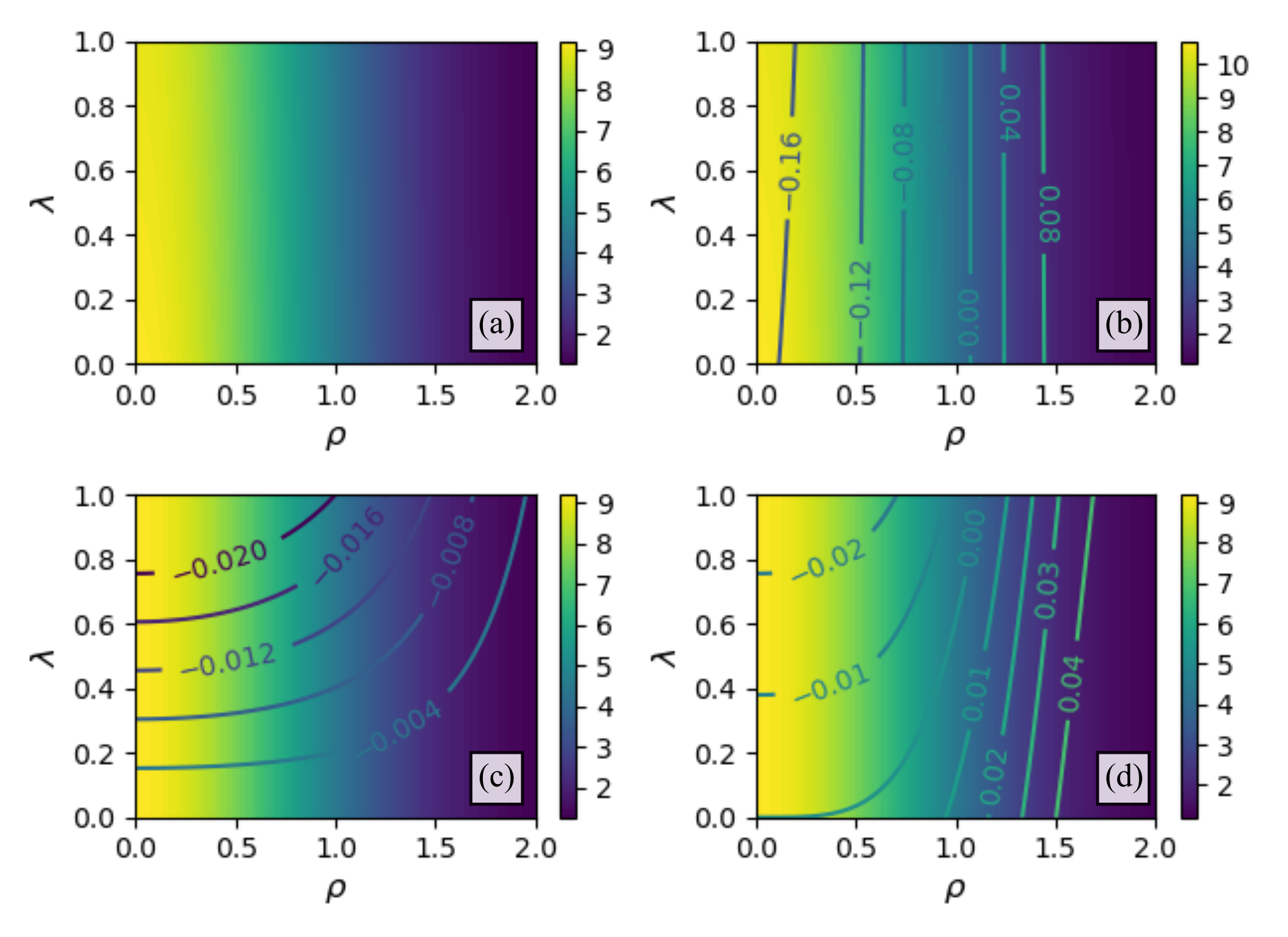}
	\caption{The temperature distribution in the sample obtained at $t = t_p = 2.95$ ms with (\textit{a}) the exact expression (\ref{expr:T_final_norm}), (\textit{b}) the normalized quasi-1D form (\ref{expr:T_final_1d_norm}), (\textit{c}) the $\lambda$-reduced 2D form (\ref{expr:T_final_norm_reduced}), (\textit{d}) the 2D power series approximation (\ref{expr:T_final_norm_reduced_F}). The tungsten properties are $\varkappa = 118$~W/(m$\cdot$K), $\rho = 19.079\times10^3$~kg/m$^3$, $C_p = 144$~J/(kg$\cdot$K). The laser spot radius $r_0$ is reduced to 0.9 mm, compared to the case shown in Fig. \ref{fig:temp_fields_Zlobinskii_case}. \label{fig:temp_fields_Zlobinskii_case_r0_0.9mm}}
\end{figure}

To further assess the accuracy of the approximations, we reduce the laser spot radius $r_0$ by around 30\%, to $r_0 = 0.9$ mm. Such a spot radius corresponds to the spot cross-section $S = 2.5$ mm$^2$. The temperature profiles, found in this case by using the previous set of laser and material properties, are shown in Fig. \ref{fig:temp_fields_Zlobinskii_case_r0_0.9mm}. It is seen that the error in the temperature profile found in the quasi-one-dimensional approximation increases up to $\approx 16$\%. Herein $\delta{T}$ isolines in the one-dimensional case extend up to the sample surface. The two-dimensional depth-reduced distributions show the error not exceeding 2\% in the hot central area of the laser spot.

\begin{figure}[H]
	\includegraphics{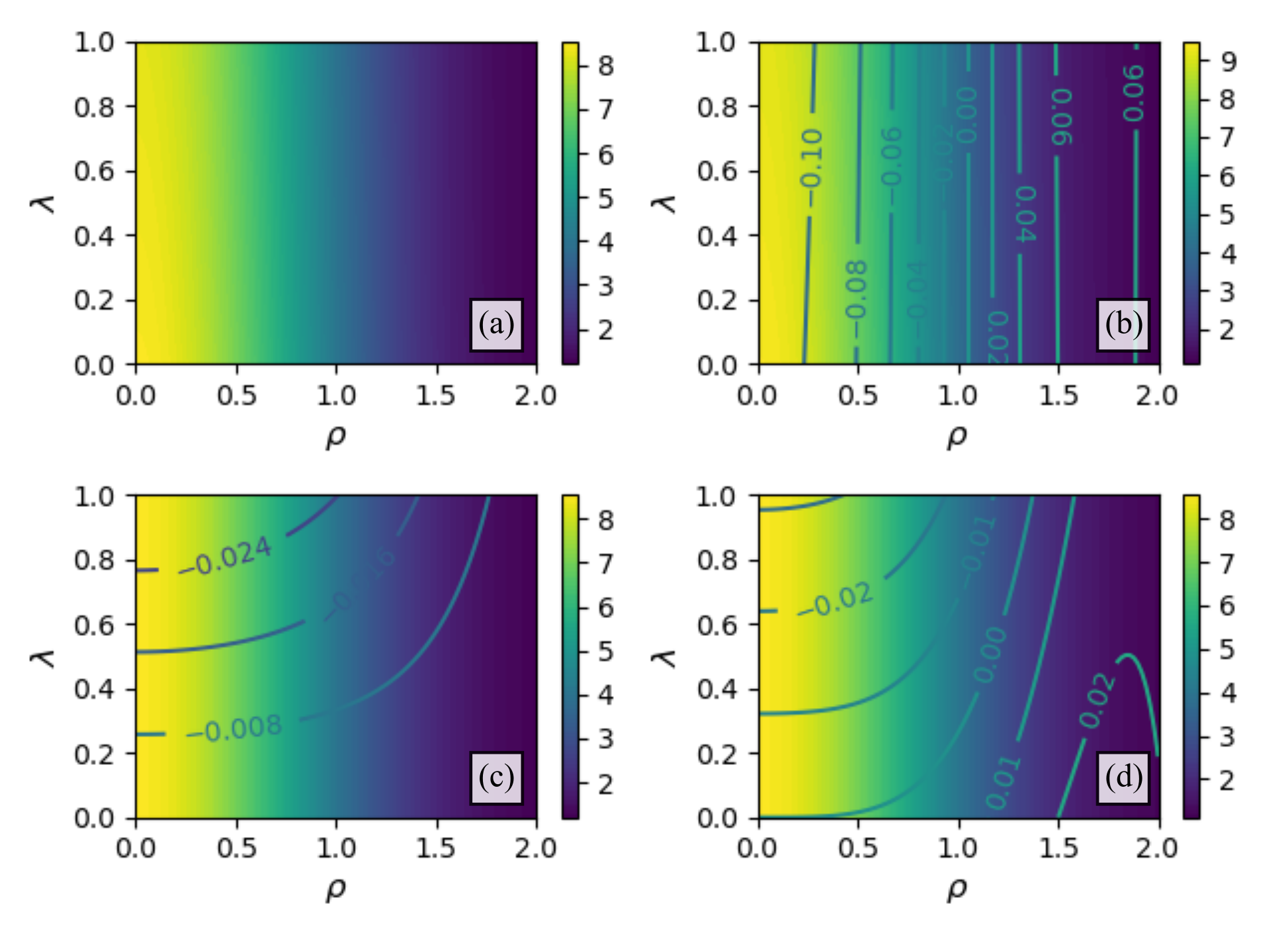}
	\caption{The temperature distribution in the sample obtained at $t = t_p = 2.95$ ms with (\textit{a}) the exact expression (\ref{expr:T_final_norm}), (\textit{b}) the normalized quasi-1D form (\ref{expr:T_final_1d_norm}), (\textit{c}) the $\lambda$-reduced 2D form (\ref{expr:T_final_norm_reduced}), (\textit{d}) the 2D power series approximation (\ref{expr:T_final_norm_reduced_F}). The tungsten properties are taken at the reference temperature of 3000 K, $\varkappa = 107.5$~W/(m$\cdot$K), $\rho = 18.22\times10^3$~kg/m$^3$, $C_p = 217.8$~J/(kg$\cdot$K).\cite{zinoviev1989} The laser spot radius is $r_0 = 0.9$ mm. \label{fig:temp_fields_T3000K_case}}
\end{figure}

The distributions shown in Fig. \ref{fig:temp_fields_Zlobinskii_case} were obtained for the tungsten thermal properties defined at the temperature 1000 K.\cite{zinoviev1989} Let us now compare the temperature distributions, if we define the tungsten thermal properties at the temperature corresponding to the maximum sample temperature in the center of the laser spot, $\approx 3000$ K. In this case,\cite{zinoviev1989} $\varkappa = 107.5$~W/(m$\cdot$K), $\rho = 18.22\times10^3$~kg/m$^3$, $C_p = 217.8$~J/(kg$\cdot$K). The re-calculated temperature profiles are shown in Fig. \ref{fig:temp_fields_T3000K_case}. The radius spot radius is kept at $r_0 = 0.9$ mm.
As seen, in the 1D case the relative error $\delta{T}$ becomes smaller, compared to the case shown in Fig. \ref{fig:temp_fields_Zlobinskii_case_r0_0.9mm}, however it reaches $\gtrsim 10$\% in the central area of the spot. The reduction of the relative error is related to the decrease of the temperature diffusivity coefficient $a^2$, associated with an almost two-fold increase of $C_p$ and a moderate reduction of $\varkappa$ and $\rho$ not exceeding 10\%.

\begin{figure}[H]
	\includegraphics{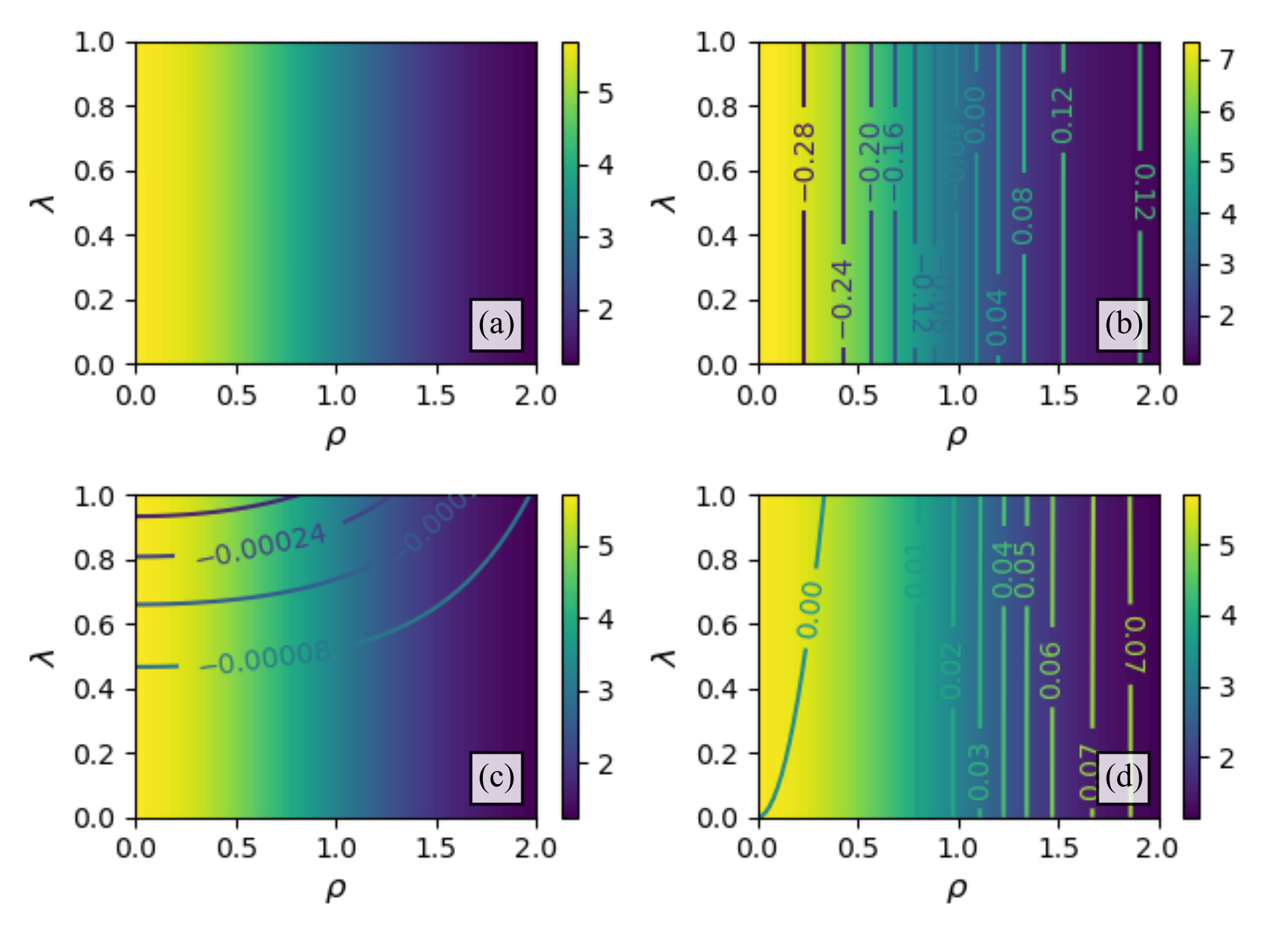}
	\caption{The temperature distribution in the sample obtained at $t = 3.5$ ms with (\textit{a}) the exact expression (\ref{expr:T_final_norm}), (\textit{b}) the normalized quasi-1D form (\ref{expr:T_final_1d_norm}), (\textit{c}) the $\lambda$-reduced 2D form (\ref{expr:T_final_norm_reduced}), (\textit{d}) the 2D power series approximation (\ref{expr:T_final_norm_reduced_F}). The tungsten properties are $\varkappa = 118$~W/(m$\cdot$K), $\rho = 19.079\times10^3$~kg/m$^3$, $C_p = 144$~J/(kg$\cdot$K). The laser spot radius is $r_0 = 0.9$ mm. \label{fig:temp_fields_Zlobinskii_case_r0_0.9mm_t3.5ms}}
\end{figure}

Since the applicability limit for the one-dimensional approximation is $r_0 \gg l_h = 2a\sqrt{t}$, we can expect the increase of the relative error $\delta{T}$ for the 1D case as time progresses beyond the pulse duration. Fig. \ref{fig:temp_fields_Zlobinskii_case_r0_0.9mm_t3.5ms} shows the temperature profiles obtained in four different approximations at $t = 3.5$ ms. For the 1D case the relative error becomes larger, compared to the previously shown data, reaching almost 30\% in the central laser spot, whereas the two-dimensional approximations show the relative error not exceeding 2\% in hot central spot.

\begin{figure}[H]
	\includegraphics{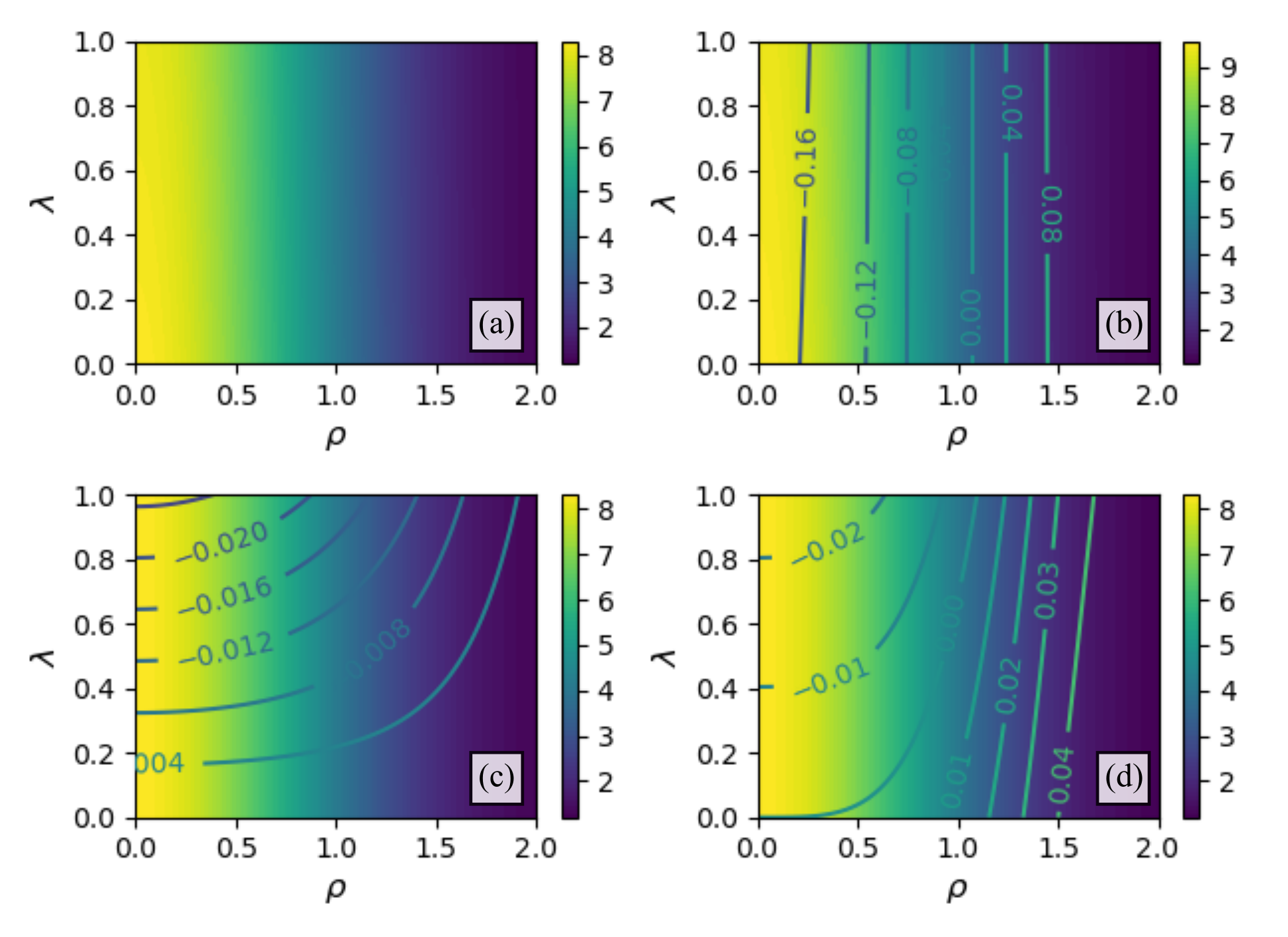}
	\caption{The temperature distribution in the sample obtained at $t = 3.5$ ms with (\textit{a}) the exact expression (\ref{expr:T_final_norm}), (\textit{b}) the normalized quasi-1D form (\ref{expr:T_final_1d_norm}), (\textit{c}) the $\lambda$-reduced 2D form (\ref{expr:T_final_norm_reduced}), (\textit{d}) the 2D power series approximation (\ref{expr:T_final_norm_reduced_F}). The tungsten properties are $\varkappa = 118$~W/(m$\cdot$K), $\rho = 19.079\times10^3$~kg/m$^3$, $C_p = 144$~J/(kg$\cdot$K). The laser spot radius is $r_0 = 0.9$ mm, the attenuation parameter $\Delta = 0.15$. \label{fig:temp_fields_Zlobinskii_case_r0_0.9mm_delta0.15}}
\end{figure}

We now show the temperature distributions for the case shown in Fig. \ref{fig:temp_fields_Zlobinskii_case_r0_0.9mm}, by introducing the non-vanishing attenuation parameter for the laser pulse, $\Delta = 0.15$. The calculation results are shown in Fig. \ref{fig:temp_fields_Zlobinskii_case_r0_0.9mm_delta0.15}. It is seen that the overall sample temperature is decreased, however the relative error distributions remain similar to those shown in Fig. \ref{fig:temp_fields_Zlobinskii_case_r0_0.9mm_t3.5ms}, with $\delta{T}$ reaching 16\% in the quasi-one-dimensional approximation, and remaining below 2\% for the two-dimensional profiles in the hot central spot.

\begin{figure}[H]
	\includegraphics{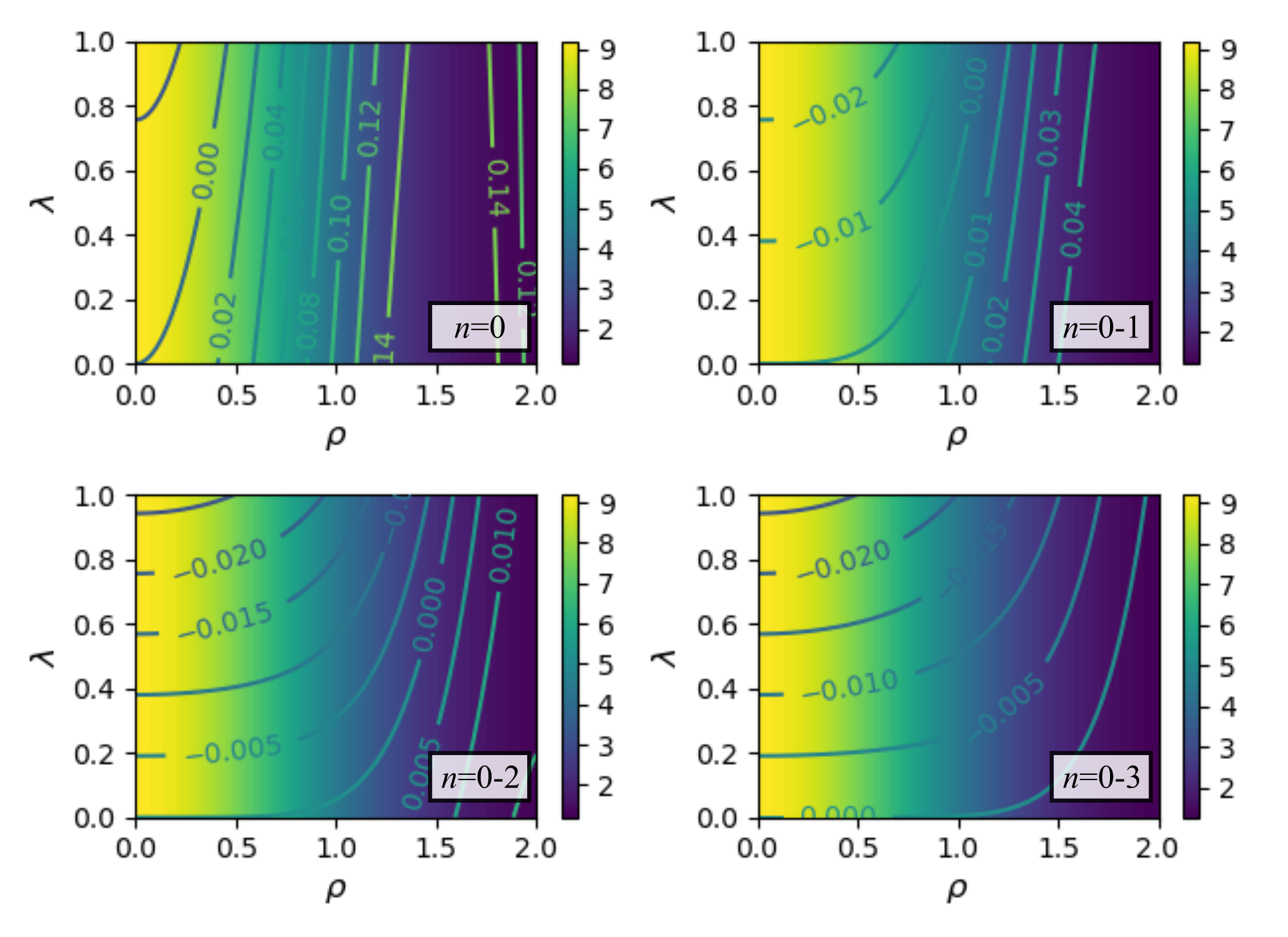}
	\caption{The temperature distribution in the sample obtained with polynomial approximation (\ref{expr:T_final_norm_reduced_F}) depending on the polynomial order $n$. \label{fig:temp_fields_Zlobinskii_case_r0_0.9mm_polynomials}}
\end{figure}

We now demonstrate how taking additional polynomials in the expression (\ref{expr:T_final_norm_reduced_F}) affect the relative error $\delta{T}$ of the temperature profile reconstruction. We use the set of parameters: $I_m = 850$ MW/m$^2$, $t = t_p = 2.95$ ms,  $r_0 = 1.3$ mm, $\Delta = 0$, $\varkappa = 118$~W/(m$\cdot$K), $\rho = 19.079\times10^3$~kg/m$^3$, $C_p = 144$~J/(kg$\cdot$K). The results are shown in Fig. \ref{fig:temp_fields_Zlobinskii_case_r0_0.9mm_polynomials}. As seen, even using only the first two polynomials, $A_0(\theta)$ and $B_0(\theta)$, yield the temperature profile deviating, to the smaller side, from the exact solution by no more than 2\% in the central laser spot. Predictably, for large values of $\rho$ the error $\delta{T}$ grows, however with the increase of $n$ it diminishes. Herein, in case of the two-term expansion ($n=0,1$) in (\ref{expr:T_final_norm_reduced_F}), the error does not exceed 4\% within the whole range of $r \leq r_0$, i.e. within the whole region of laser heating, where the gas desorption primarily occurs.

\begin{figure}[H]
	\includegraphics{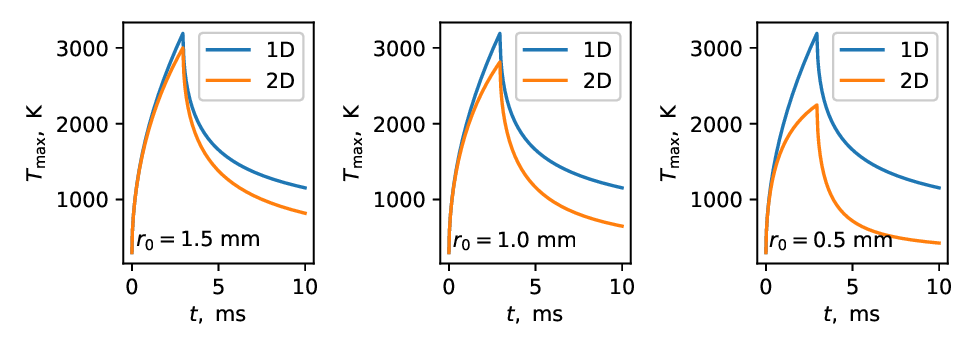}
	\caption{The temporal dependencies of the maximum sample temperature at the surface obtained by using the one- and two-dimensional approximations for different values of the laser spot radius $r_0$.\label{fig:max_temp}}
\end{figure}

The temporal dependencies of the maximum sample temperature at the surface, found by using one- and two-dimensional approximations, are shown in Fig. \ref{fig:max_temp}. For all cases, we choose tungsten, its thermal properties are taken at $T = 1000$ K, so that $\varkappa = 118$~W/(m$\cdot$K), $\rho = 19.079\times10^3$~kg/m$^3$, $C_p = 144$~J/(kg$\cdot$K). The laser spot radius takes values $r_0 = 0.5, 1.0, 1.5$ mm. As seen, for a large-radius spot, $r_0 = 1.5$ mm, the differences between the 1D and 2D dependencies are rather small during the heating phase, reaching 6.4\%. During the cooling stage ($t > t_p$) the relative error grows and reaches 41\%. With the decrease of the laser spot radius, the differences between two curves increase. For $r_0 = 0.5$ mm, the maximum difference between $T_{\max}$ in 1D and 2D cases reaches 42\% during the heating stage and is on the order of 171\% during the sample cooling. Thus it can be envisaged that with the decrease of $r_0$ the discrepancy between the one- and two-dimensional temperature profiles will impact the gas desorption from the sample. The values of the maximum surface temperature found at $t = t_p$ for different values of the laser spot radius $r_0$ and the pulse intensity attenuation rate $\Delta$ are summarized in Table \ref{tab:Tmax_Delta}.
\begin{table}[H]
	\caption{The maximal values of the surface temperature depending on the laser spot radius $r_0$ and the pulse intensity attenuation rate $\Delta$.  \label{tab:Tmax_Delta}}
	\begin{ruledtabular}
		\begin{tabular}{ccccccc}
			\multirow{3}{*}{$\Delta$}	&	\multicolumn{6}{c}{$T_{\max}$, kK} \protect\\
			\cmidrule{2-7}
			&	\multicolumn{2}{c}{$r_0 = 0.5$ mm} &	\multicolumn{2}{c}{$r_0 = 1.0$ mm} &	\multicolumn{2}{c}{$r_0 = 1.5$ mm} \protect\\
			\cmidrule{2-3} \cmidrule{4-5} \cmidrule{6-7}
			    & 1D & 2D & 1D & 2D & 1D & 2D \protect\\ \hline
			0   & 2.99 & 2.27 & 2.99 & 2.73 & 2.99 & 2.86 \\
			0.1 & 2.81 & 2.12 & 2.81 & 2.56 & 2.81 & 2.69 \\
			0.2 & 2.63 & 1.97 & 2.63 & 2.39 & 2.63 & 2.51 \\
			0.3 & 2.45 & 1.83 & 2.45 & 2.22 & 2.45 & 2.34 
		\end{tabular}
	\end{ruledtabular}
\end{table}

\section{SENSITIVITY ANALYSIS\label{sec:sensitivity_analysis}}
The demonstrated results were based on using the linear heat equation. However, the thermal properties of the sample depend on the temperature and the sample composition (e.g. metallic films re-deposited on the samples can have thermal properties deviating from the pure material characteristics). In addition, the light absorption by the sample may vary depending on the surface condition, temperature, etc. All these factors can affect the reconstructed temperature distribution in the sample. For these reasons, in what follows we shall perform the sensitivity analysis of the derived results to uncertainties in the transport properties of the sample.

\begin{figure}[H]
	\centering
	\includegraphics[scale=0.8]{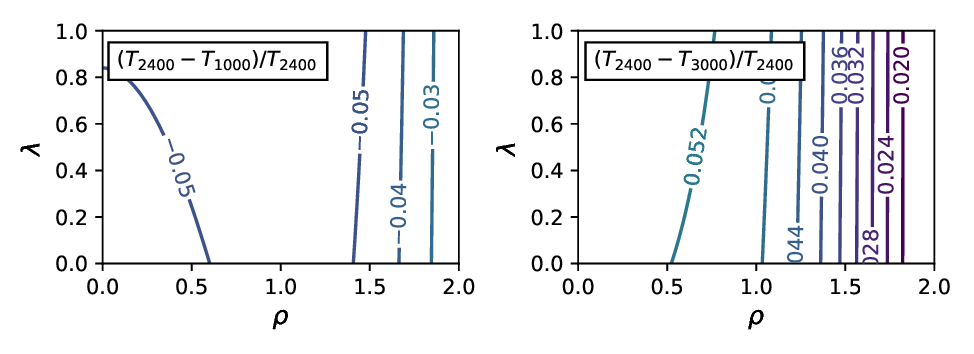}
	\caption{The relative error $\delta{T}$ distribution between the temperature fields, obtained by using the tungsten thermal properties, defined at $1000, 2400, 3000$ K. The laser spot radius is $r_0 = 1.5$ mm.\label{fig:relative_errors_r0_1.5mm}}
\end{figure}

\begin{figure}[H]
	\centering
	\includegraphics[scale=0.8]{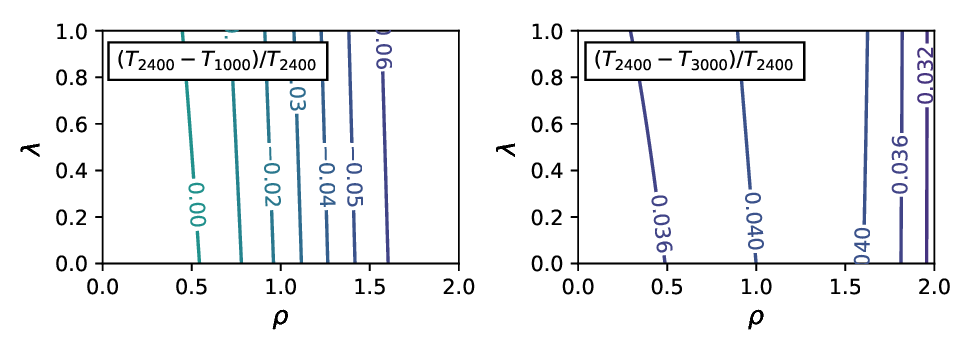}
	\caption{The relative error $\delta{T}$ distribution between the temperature fields, obtained by using the tungsten thermal properties, defined at $1000, 2400, 3000$ K. The laser spot radius is $r_0 = 0.5$ mm.\label{fig:relative_errors_r0_0.5mm}}
\end{figure}

As mentioned in Sec. \ref{sec:preliminaries}, we can expect the temperature dynamics governed by the non-linear heat equation to be bounded by the solutions obtained with the linear heat equation and a set of sample thermal properties taken at two limiting temperatures. Previously, Sec. \ref{sec:approximations}, we have seen that alterations of the material properties with the temperature can lead to noticeable variations between the obtained temperature profiles, up to 10\%, cf. the maximal temperatures in Figs. \ref{fig:temp_fields_Zlobinskii_case} and \ref{fig:temp_fields_T3000K_case}. Hence, to minimize the discrepancies between the $T$ profiles in the high- and low-temperature limits, an intermediate temperature $1000\ \mathrm{K} = T_{\min}< T_* < T_{\max} = 3000\ \mathrm{K}$ can be chosen. The direct search method of $T_*$ has shown that it is approximately equal to $T_* \approx 2400$ K. Fig. \ref{fig:relative_errors_r0_1.5mm} shows the distributions of relative errors $(T_{2400}-T_{1000})/T_{2400}$ (the left panel) and $(T_{2400}-T_{3000})/T_{3000}$ (the right panel), where the subscripts denote the temperatures, at which the values of the thermal properties were taken. The temperature profiles were obtained by using the two-term ($n=0,1$) expansion (\ref{expr:T_final_norm_reduced_F}). The laser parameters were taken identical to previous cases, herein $r_0 = 1.5$ mm, $\Delta = 0$. For a smaller spot radius, $r_0 = 0.5$ mm, the difference between the profiles becomes smaller, see Fig. \ref{fig:relative_errors_r0_0.5mm}. Values of $\varkappa$, $C_p$ and $\rho$ were taken from Ref. \onlinecite{zinoviev1989} and are summarized in Table \ref{tab:W_thermal_properties}.
\begin{table}[H]
	\caption{Data on the tungsten thermal properties.\label{tab:W_thermal_properties}}
	\begin{ruledtabular}
		\begin{tabular}{cccc}
			T, K						&	1000	&	2400	&	3000	\\\hline
			$\varkappa$, W/(m$\cdot$K) 	&	118		&	106.8	&	107.5	\\
			$C_p$, J/(kg$\cdot$K)	 	&	144.5	&	189.1	&	217.8	\\
			$\rho$, $10^3$ kg/m$^3$	 	&	19.1	&	18.52	&	18.22	
		\end{tabular}
	\end{ruledtabular}
\end{table}

As seen in the figures, the relative error between the pairs of temperature fields does not exceed 5\% in the central region of the laser spot, where the temperature takes the highest values and the desorption rates are expected to take the highest values. This error bar complies with the one introduced in Sec. \ref{sec:preliminaries}, indicating that the heat transfer in tungsten can be evaluated by taking the thermal properties of the metal at the constant temperature of 2400 K.

Another source of uncertainty is the dynamics of light absorption by the target material. The light emitted by the laser and hitting the sample is partially reflected from the surface. The reflectivity depends on the target temperature,\cite{ujihara1972reflectivity} surface roughness,\cite{cao2019surface} and other factors, variations of which are hard to be experimentally quantified. Nevertheless, this uncertainty can be reduced by using data on the maximum surface temperature measured in the experiment. The relation (\ref{expr:max_temp}) coupled with the set of the laser pulse and material thermal properties, can be used to recover the maximum light intensity $I_m$, which, within the presented analysis, incorporates the contribution from the laser light reflection from the sample surface.

Finally, another source of uncertainty is the depth of the laser light penetration into the sample surface. The extinction coefficient, $\alpha$, depends not only on the wavelength of the light, but also on the material properties and surface condition (roughness, presence of impurities, etc.). Hence, variations in $\alpha$ may be expected during an actual laser pulse. However, these differences must play a role on short timescales, when the thermal conduction length $l_h = 2a\sqrt{t}$ is smaller or comparable to the attenuation length $1/\alpha$, i.e. $2a\sqrt{t}\alpha \lesssim 1$ or $t \lesssim 1/(4a^2 \alpha)$. For tungsten at $T = 1000$ K, we have $a^2 = 4.3\times10^{-5}$ m$^2$/s, $1/\alpha \approx 8$ nm, and $t \sim 50\ \mu\mathrm{s} \ll t_p$. Thus, for the laser pulses characteristic for the LID technique, the extinction length for the laser radiation can be taken as infinitely small.

\section{IMPACT OF THE TEMPERATURE DISTRIBUTION ON GAS DESORPTION FROM TARGET\label{sec:gas_desorption}}
As a final test for the demonstrated results, we assess how different approximations for the sample temperature affect the particle desorption dynamics. To describe the particle transport in the sample, we use the set of reaction-diffusion equations for the solute, $u$, and trapped, $y$, tritium concentration
\begin{align}
	\ddtp{u} &= \nabla\left[D_0\exp\left(-\frac{E_d}{kT}\right)\nabla{u}\right] - \ddtp{y}, \label{eqn:u}\\
	\ddtp{y} &= \nu\exp\left(-\frac{E_d}{kT}\right) \left[\frac{u}{u_m}(y_m - y) - y\exp\left(-\frac{E_b}{kT}\right)\right], \label{eqn:y}
\end{align}
where $D_0 = 4\times10^{-7}$ m$^2$/s is the diffusion pre-exponent, $E_d$ is the diffusion activation energy, $\nu=10^{13}$ s$^{-1}$ is the Debye frequency, $u_m$ and $y_m$ are the maximum solute and trapped tritium concentrations, respectively, $E_b$ is the trap binding energy. The temperature $T$ is evaluated by using either the quasi-one-dimensional (\ref{expr:T_final_1d_norm}) or the two-dimensional (\ref{expr:T_final_norm_reduced_F}) approximations.

The diffusion is considered in the semi-infinite sample, with the material-vacuum boundary located at $z = 0$ (the reference system for the problem is the same as for the heat transport problem). The sample thickness is $L$. The boundary conditions imposed on $u$ are
\begin{align}
	u(r,0,t) = 0, \qquad \left.\frac{\partial{u}}{\partial{r}}\right|_{r = R} = \left.\frac{\partial{u}}{\partial{r}}\right|_{r = 0} = \left.\frac{\partial{u}}{\partial{z}}\right|_{z = L} = 0,
\end{align}
where $R$ is the radial size of the considered domain. The initial conditions are
\begin{align}
	u(r,z,0) = 0, \qquad y(r,z,0) = y_m, \label{eqn:ic}
\end{align}
corresponding to the film of the thickness $L$, completely filled with tritium in the traps.

The system (\ref{eqn:u})-(\ref{eqn:ic}) was solved numerically with the aid of the BOUT++ framework.\cite{dudson2009bout++,dudson2015bout++} The diffusion activation energy, $E_d$, the trap binding energy $E_b$, the maximum solute, $u_m$, and trapped, $y_m$, tritium concentrations, the laser intensity $I_m$, spot radius $r_0$, pulse duration $t_p$, pulse attenuation factor $\Delta$, the sample thickness $L$, and the radial domain size $R$ were as follows: $E_d = 0.39$ eV, $E_b = 0.5, 2.0$ eV, $u_m = 6.31\times10^{28}$ part./m$^3$, $y_m = 0.1 u_m$, $I_m = 850$ MW/m$^2$, $r_0 = 0.5, 1.0, 1.5$ mm, $t_p = 3$ ms, $\Delta = 0.0$, $L = 10, 30\ \mu\mathrm{m}$, $R = 3r_0$. The laser spot radius and trapping energies were taken so as to consider six different cases of gas release from the sample having defects with the low/high trapping energies under laser irradiation of the spot with the small/medium/large radius. The sample thickness was set to $L = 10\ \mu\mathrm{m}$, when $E_b = 2$ eV. For defects having $E_b = 0.5$ eV, we set $L = 30\ \mu\mathrm{m}$ to prevent the complete de-gassing of the sample. The numerical grid for simulations had the resolution $N_r \times N_z = 120 \times 120$ points.

The total number of particles desorbed from the sample under laser irradiation, $N_{des}$, found in simulations, are summarized in Table \ref{tab:total_N_desorbed}.
\begin{table}[H]
	\caption{The total number of particles desorbed from the tungsten sample under laser irradiation, found in the quasi-one- and two-dimensional approximations. The relative error $\delta{N}$ is defined with respect to the two-dimensional case.\label{tab:total_N_desorbed}}
	\begin{ruledtabular}
		\begin{tabular}{cccccccccc}
			\multirow{3}{*}{$E_{t}$, eV}	&	\multicolumn{3}{c}{$r_0 = 0.5$ mm} &	\multicolumn{3}{c}{$r_0 = 1.0$ mm} & \multicolumn{3}{c}{$r_0 = 1.5$ mm}  \\
			\cmidrule{2-4}\cmidrule{5-7}\cmidrule{8-10}
			&	\multicolumn{2}{c}{$N_{des}$, part.}	& \multirow{2}{*}{$\delta{N}$,\%} &
			\multicolumn{2}{c}{$N_{des}$, part.}	& \multirow{2}{*}{$\delta{N}$,\%} &	
			\multicolumn{2}{c}{$N_{des}$, part.}	& \multirow{2}{*}{$\delta{N}$,\%} \\
			\cmidrule{2-3}\cmidrule{5-6}\cmidrule{8-9}
			&	1D		&		2D		&			& 	1D		&		2D		&	 &	1D		&		2D &\\
			\hline
			0.5		&	$3.62\times10^{16}$		&		$2.18\times10^{16}$		&	66
					&	$1.44\times10^{17}$		&		$1.24\times10^{17}$		&	16
					&	$3.26\times10^{17}$		&		$3.38\times10^{17}$		& -3.6\\
			2.0		&	$1.44\times10^{15}$		&		$4.09\times10^{14}$		&	251	
					&	$5.84\times10^{15}$		&		$4.12\times10^{15}$		&	42
					&	$1.31\times10^{16}$		&		$1.14\times10^{16}$		& 15
		\end{tabular}
	\end{ruledtabular}
\end{table}
From the data shown, it is seen that in case of the large laser spot, $r_0 = 1.5$ mm, the total number of desorbed particles, obtained in the quasi-one- and two-dimensional approximations, do not strongly differ from each other. The relative error between a pair of readings is, expectedly, the largest ($\approx 15\%$) for traps with the large trapping energy, $E_b = 2$ eV. For low-energy traps, the difference between a pair of values $N_{des}$ does not exceed 4\%.

For the laser beam with the medium values of the laser spot radius, $r_0 = 1.0$ mm, the discrepancies between values of $N_{des}$ grow by approximately 3-4 times. For low-energy traps, the deviation between two limits remains rather small, not exceeding $\approx 16\%$.  The number of particles released from the traps with the large binding energy differ by 42\%.

The situation completely changes in case of the small-radius laser spot, $r_0 = 0.5$ mm. In this case, even for low-energy traps, $E_b = 0.5$ eV, the difference between two values of $N_{des}$ in the one- and two-dimensional approximations reaches 66 \%. For high-energy traps, the difference increases up to 251 \%, i.e. to an almost four-fold ratio.

Thus, from the demonstrated results, it is clearly seen that one-dimensional approximation is applicable only in case of the sample irradiation by the laser beam with a sufficiently large radius. For small-radius beams, the application of the one-dimensional approximation can lead to large errors in the temperature fields, resulting in significant errors in the number of particles desorbed from the sample under irradiation, compared to the full two-dimensional approach. As a result, the dimensional effects must be taken into account to accurately resolve the particle desorption dynamics under sample irradiation by focused laser beams.

\section{CONCLUSIONS\label{sec:conclusions}}
In this study, we have analyzed the conditions, under which the temperature field dynamics in a sample under laser pulse irradiation characteristic for the LID diagnostic can be treated either within quasi-one-dimensional or two-dimensional approximations. 

The qualitative consideration of the heat propagation in the sample was given. The analysis showed that if the laser spot radius is comparable to the heat diffusion length on the laser pulse timescale, the heat transport in the sample should be treated by taking the dimensional effects into account. 

The rigorous mathematical analysis of the problem allowed obtaining both one- and two-dimensional representations of the temperature profile resolved in time. By using these relations, it was demonstrated that depending on the laser spot radius the temperature distributions in the sample taken in the two opposite limits may differ from each other. Herein the one-dimensional approximation for the temperature profile overestimates $T$, and the discrepancies with the two-dimensional limit grow as the spot radius is diminished. 

For the test problem of tungsten heated by a laser pulse,\cite{zlobinski2011laser} it was found that the differences in the one-dimensional approximation can reach $\sim 10\%$ in the hot central spot area all the way up to the sample surface at the end of the laser pulse, increasing over time as heating is turned off and the sample cools down. For the different forms of the two-dimensional temperature profile the relative error with respect to the exact distribution of the sample temperature did not exceed 2\%. Herein the error vanished in the near-surface region of the sample. In the beam focal point on the sample surface, the deviations between the temperature temporal profiles obtained in the two limits could reach as high as $\gtrsim 100\%$ as the laser beam spot radius was diminished from $r_0 = 1.5$ mm to $r_0 = 0.5$ mm.

The sensitivity analysis of the derived relations to experimental uncertainties in the values of the material thermal properties, radiation absorption and reflection parameters was performed. It was shown that for modeling temperature dynamics of the irradiated sample during LID the values of the material thermal properties defined at some intermediate temperature between the maximum and minimum temperatures of the target can be chosen. For tungsten, this temperature was estimated as $\approx 2400$ K. Herein the obtained temperature profile deviated from the distributions, obtained by using the tungsten thermal properties at the limiting temperatures of 1000 K and 3000 K, did not exceed 5\%.

Finally, the derived relations were applied to modeling tritium desorption dynamics from a sample, characterized by two different trap binding energies, $E_b = 0.5, 2$ eV. It was demonstrated that in case of the sample irradiation by a laser beam with the large radius $r_0 = 1.5$ mm yielded the comparable amounts of tritium desorbed from the sample. The deviation between the found values of $N_{des}$ did not exceed 15\%. With the decrease of the laser spot radius to $r_0 = 0.5$ mm, the discrepancies between the values of the total number of particles desorbed from the sample increased. For traps with low binding energies, $E_b = 0.5$ eV, the difference between two measures reached $\approx 66\%$. In case of traps with a large binding energy, $E_b = 2$ eV, the total number of desorbed particles in the quasi-one-dimensional approximation was higher than its two-dimensional counterpart by almost 3.5 times, indicating the importance of accurately resolving the temperature profiles in sample irradiated by focused laser beams. Herein it can be envisaged that in the rigorous one-dimensional approximation the total number of particles desorbed from the sample will be overestimated even further compared to the  quasi-one-dimensional approach due to the estimation of the desorption rates at distances $r > 0$ with the temperature profile in the center of the laser spot. Thus, accounting for the dimensional effects in the description of the particle release from the sample under pulsed laser irradiation is of high importance for the successful interpretation of the LID diagnostic data.

\section*{ACKNOWLEDGMENTS}
The work was supported by the Ministry of Science and Higher Education of the Russian Federation (project No. 0723-2020-0043).

\section*{AUTHOR DECLARATIONS}
\subsection*{Conflict of Interest}
The authors have no conflicts to disclose.

\section*{DATA AVAILABILITY}
The data that support the findings of this study are available from the corresponding author upon reasonable request.

\section*{APPENDIX\label{sec:appendix}}	
To evaluate the limit: $\lim\limits_{1/\alpha \rightarrow 0}f(z,\alpha)$, where $f(z,\alpha) = \alpha\exp\left(-\alpha z\right)$ -- we introduce the auxiliary variable, $l = 1/\alpha$, and function $g(z,l) = \exp(-|z|/l)/(2l)$. The integral $\int_{-\infty}^{+\infty}g(z,l)\dif{z} = 1$. Hence, $\lim\limits_{l \rightarrow 0}g(z,l) = \delta(z)$ in the sense of a weak limit. As a result, $f(z,\alpha) = \alpha\exp\left(-\alpha z\right) \xrightarrow{\alpha \rightarrow 0} 2\delta(z)$, when $z > 0$, so that $\int_{0}^{+\infty}dzA(z)f(z,\alpha) \xrightarrow{\alpha \rightarrow 0} A(0)$.

\bibliographystyle{aipnum4-1}
\bibliography{target-temp-dyn-LID}
\end{document}